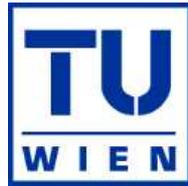

Technische Universität Wien

DIPLOMARBEIT

# Electrodynamics on the Möbius Strip

Ausgeführt am Institut für

Theoretische Physik

der Technischen Universität Wien

unter der Anleitung von

Ao. Univ. Prof. Dr. Karl Svozil

durch

Gernot Pauschenwein

Ing. J. Raab–Gasse 10, 7203 WIESEN

25. 08. 2004                    Gernot Pauschenwein e.h.

II

# Preface

The idea of taking a closer look at the Möbius strip in electrodynamics originated from my colleague and diploma–guide Karl Svozil, who I have already worked together over years, giving lectures about methods of theoretical physics.

I thank him for letting me spread the diploma work in whatever way my ideas led me and on the other hand leading me to a meaningful end. This is what emerged from our collaboration.

IV

# Contents

# Chapter 1

# Introduction

In this diploma work electrostatics and electrodynamics on two–dimensional structures is performed.

As a consequence of the original idea, which was to find out whether a Möbius strip could be used to build some new motor or other device, the Möbius strip will be taken as an example–structure throughout this paper. It is used to show the way the programmed *Mathematica*–packages work. Therefore the Möbius strip and its parametrisation are introduced in the second chapter.

In the third chapter, how to calculate the total charge of a two–dimensional charged structure and all about the *Mathematica*–package `AreaPotential.m` is explained. With help of this, the electrostatic (and even any other kind of) potential of arbitrary charged surfaces can be calculated and visualized.

If a conductor is present one cannot choose how a given charge spreads over the surface of the conductor. To calculate the partition of a given charge `AreaConductor.m` devides the total charge into an arbitrary number of point–charges and searches an equilibrium of their positions while being restricted to the conductor–surface. This numerical process is exported from *Mathematica* to C. All this is done in detail in chapter four.

For applications the radiation of a rotating structure is of interest. This is discussed in the fifth chapter up to electric quadrupole–radiation.

All the mentioned files (the *Mathematica*–packages `AreaPotential.m` and `AreaConductor.m`, the C–files `potential.template` and `conductor.template` and the corresponding `.tm`–files) are given in the appendix or can be downloaded at http://k3.itp.tuwien.ac.at/~gernot.

The *Mathematica* commands used throughout this work can also be found in the notebooks `session1.nb` (potential of given charge distributions, chapter 3), `session2.nb` (distribution of charges on a conductor and the resulting potential, chapter 4), `session3.nb` and `session4.nb` (radiation of both types under rotation, chapter 5).

# Chapter 2

# The Möbius Strip

In this paper the well known parametrisation of the Möbius strip in 3 dimensions (only basic analysis will be necessary, any lecture notes about analysis for physicians will do, e.g. [6]) is used:

$$
\mathbf{m}(\varphi, t) = \begin{pmatrix} \cos(\varphi) + t \cdot \cos\left(\frac{\varphi}{2}\right) \cdot \cos(\varphi) \\ \sin(\varphi) + t \cdot \cos\left(\frac{\varphi}{2}\right) \cdot \sin(\varphi) \\ t \cdot \sin\left(\frac{\varphi}{2}\right) \end{pmatrix}, \tag{2.1}
$$

$$
\mathcal{M} = \left\{ \mathbf{x} \epsilon \mathbb{R}^3 \, \middle| \, \mathbf{x} = \mathbf{m}(\varphi, t), 0 \le \varphi < 2\pi, -l \le t \le l \right\}. \tag{2.2}
$$

For the following calculations one will need the normal vector on every point of the Möbius strip in this parametrisation:

$$
\mathbf{n} = \mathbf{m}_\varphi \times \mathbf{m}_t =
$$

$$
= \begin{pmatrix} -\frac{1}{2} \cdot \sin\left(\frac{\varphi}{2}\right) \cdot \left(4 \cdot \cos\left(\frac{\varphi}{2}\right) + t \cdot (2 + 3\cos(\varphi))\right) \\ \frac{1}{4} \cdot t \cdot \cos\left(\frac{\varphi}{2}\right) + \cos(\varphi) + \frac{3}{4} \cdot t \cdot \cos\left(\frac{3\varphi}{2}\right) \\ \frac{1}{2} \cdot t \cdot \cos\left(\frac{\varphi}{2}\right) \end{pmatrix} \times \begin{pmatrix} \cos\left(\frac{\varphi}{2}\right)\cos(\varphi) \\ \cos\left(\frac{\varphi}{2}\right)\sin(\varphi) \\ \sin\left(\frac{\varphi}{2}\right) \end{pmatrix} =
$$

$$
= \begin{pmatrix} \frac{1}{2} \cdot \sin\left(\frac{\varphi}{2}\right) \cdot \left(-t\cos\left(\frac{\varphi}{2}\right) + 2\cos(\varphi) + t \cdot \cos\left(\frac{3\varphi}{2}\right)\right) \\ \frac{1}{4} \cdot \left(t + 2\cos\left(\frac{\varphi}{2}\right) + 2t \cdot \cos(\varphi) - 2\cos\left(\frac{3\varphi}{2}\right) - t \cdot \cos(2\varphi)\right) \\ -\cos\left(\frac{\varphi}{2}\right) \cdot \left(1 + t \cdot \cos\left(\frac{\varphi}{2}\right)\right) \end{pmatrix}
$$

$$\tag{2.3}$$

and its length for the infinitesimal area-element $dA$.

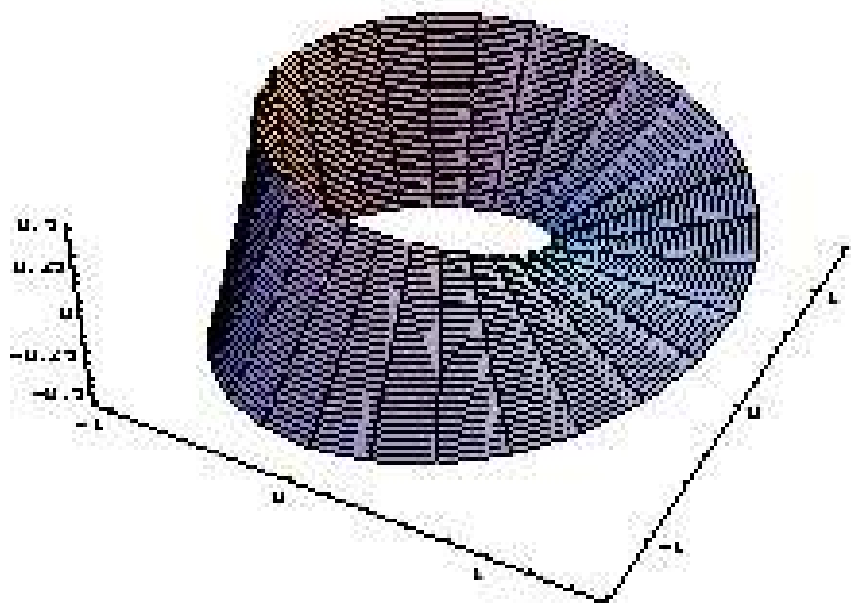

Figure 2.1: Möbius strip for $l = 1/2$.

# Chapter 3

# Static charged Areas

The point of interest is not only how to calculate and visualise the electrostatic potential of the desired charged areas, but also to compare their potential with that of well known structures. Therefore it is important to know the total charge of a given area–charge–density.

## 3.1 Total Charge

The general formula for the total charge of an area is simply the integral of the scalar area–charge–density $\sigma$ over the area $\mathcal{A}$ (see [1, 5]):

$$Q_{\text{tot}} = \int_{\mathcal{A}} \sigma \, dA \; . \tag{3.1}$$

Now one takes the Möbius strip as an example. To calculate the integral we take the parametrisation of the previous chapter for the area of the Möbius strip:

$$
\begin{aligned}
Q(l) &= \int_{strip} \sigma \, dO = \int_0^{2\pi} \int_{-l}^{l} \sigma |\mathbf{n}| \, dt \, d\varphi = \\
&= \frac{1}{2} \int_0^{2\pi} \int_{-l}^{l} \sigma \sqrt{4 + 3t^2 + 8t \cdot \cos\left(\frac{\varphi}{2}\right) + 2t^2 \cdot \cos(\varphi)} \, dt \, d\varphi \; . \tag{3.2}
\end{aligned}
$$

Unfortunately, even if the charge is equally distributed over the area ($\sigma = $ const.), only the $t$–integration is (trusting *Mathematica*) analytically possible and results in a rather nasty thing:

```
In[1]:= Integrate[√((1 + t Cos[f/2])² + 1/4 t²), {f, 0, 2π}, {t, -1, 1}]
```

$$Out[1]= \int_0^{2\pi} \frac{1}{4\,(3+2\,Cos[f])^{3/2}}$$

$$\left(3\,1\,\sqrt{(3+2\,Cos[f])\,\left(4+3\,1^2-8\,1\,Cos\left[\frac{f}{2}\right]+2\,1^2\,Cos[f]\right)}+\right.$$

$$3\,1\,\sqrt{(3+2\,Cos[f])\,\left(4+3\,1^2+8\,1\,Cos\left[\frac{f}{2}\right]+2\,1^2\,Cos[f]\right)}+$$

$$4\,Cos\left[\frac{f}{2}\right]\,\left(-\sqrt{(3+2\,Cos[f])\,\left(4+3\,1^2-8\,1\,Cos\left[\frac{f}{2}\right]+2\,1^2\,Cos[f]\right)}+\right.$$

$$\left.\sqrt{(3+2\,Cos[f])\,\left(4+3\,1^2+8\,1\,Cos\left[\frac{f}{2}\right]+2\,1^2\,Cos[f]\right)}\right)+$$

$$2\,1\,Cos[f]\,\left(\sqrt{(3+2\,Cos[f])\,\left(4+3\,1^2-8\,1\,Cos\left[\frac{f}{2}\right]+2\,1^2\,Cos[f]\right)}+\right.$$

$$\left.\sqrt{(3+2\,Cos[f])\,\left(4+3\,1^2+8\,1\,Cos\left[\frac{f}{2}\right]+2\,1^2\,Cos[f]\right)}\right)-$$

$$8\,Log[2]-4\,Log\left[1-\frac{2\,Cos[\frac{f}{2}]}{\sqrt{3+2\,Cos[f]}}\right]-4\,Log\left[1+\frac{2\,Cos[\frac{f}{2}]}{\sqrt{3+2\,Cos[f]}}\right]+$$

$$4\,Log\left[\frac{1}{\sqrt{3+2\,Cos[f]}}\left(3\,1-4\,Cos\left[\frac{f}{2}\right]+2\,1\,Cos[f]+\right.\right.$$

$$\left.\left.\sqrt{(3+2\,Cos[f])\,\left(4+3\,1^2-8\,1\,Cos\left[\frac{f}{2}\right]+2\,1^2\,Cos[f]\right)}\right)\right]+$$

$$4\,Log\left[\frac{1}{\sqrt{3+2\,Cos[f]}}\left(3\,1+4\,Cos\left[\frac{f}{2}\right]+2\,1\,Cos[f]+\right.\right.$$

$$\left.\left.\left.\sqrt{(3+2\,Cos[f])\,\left(4+3\,1^2+8\,1\,Cos\left[\frac{f}{2}\right]+2\,1^2\,Cos[f]\right)}\right)\right]\right)df\ .$$

A numerical integration on the other hand is no problem. Taking the total width of the Möbius strip to be 1 ($l = 0, 5$) one gets

$$Q(0,5) = \frac{\sigma}{2}\int_0^{2\pi}\int_{-0,5}^{0,5}\sqrt{4+3t^2+8t\cdot\cos\left(\frac{\varphi}{2}\right)+2t^2\cdot\cos(\varphi)}\,dt\,d\varphi = 6,35327\cdot\sigma \tag{3.3}$$

which is, due to distortions because of the parametrisation, more than $6,28319\cdot\sigma$,

the charge of a flat strip of area $2\pi \times 1$.

## 3.2 Electrostatic Potential

The potential $\Phi(\mathbf{r})$ can be calculatet by using the Green's–function

$$G(\mathbf{r}, \mathbf{r}') = -\frac{1}{4\pi} \frac{1}{|\mathbf{r} - \mathbf{r}'|} \tag{3.4}$$

for the Laplace–equation $\triangle \Phi = -4\pi\rho$ (see again [1, 5]). Applied to an area–charge and especially to the one of the Möbius strip one gets

$$
\begin{aligned}
\Phi(x, y, z) &= \int_{strip} \frac{\sigma}{|\mathbf{r} - \mathbf{r}'|} \, dA' = \\
&= \int_0^{2\pi} \int_{-l}^{l} \sigma(t, \varphi) \sqrt{4 + 3t^2 + 8t \cdot \cos\left(\frac{\varphi}{2}\right) + 2t^2 \cdot \cos(\varphi)} \cdot \\
&\quad \cdot \frac{1}{\sqrt{\left(x - \cos(\varphi) - t \cdot \cos\left(\frac{\varphi}{2}\right)\cos(\varphi)\right)^2 + \left(y - \sin(\varphi) - t \cdot \cos\left(\frac{\varphi}{2}\right)\sin(\varphi)\right)^2 + \left(z - t \cdot \sin\left(\frac{\varphi}{2}\right)\right)^2}} \, dt \, d\varphi \, .
\end{aligned}
\tag{3.5}
$$

If $\sigma$ is constant it can be pulled out of the integral, but the remaining integration is still quite challenging. Therefore one first takes a look at the

### 3.2.1 Potential on the z–Axis

$$
\Phi(z) = \Phi(0, 0, z) = \int_{strip} \frac{\sigma}{|z\mathbf{e}_z - \mathbf{r}'|} dO' = \tag{3.6}
$$

$$
= \sigma \int_0^{2\pi} \int_{-l}^{l} \frac{\sqrt{4 + 3t^2 + 8t \cdot \cos\left(\frac{\varphi}{2}\right) + 2t^2 \cdot \cos(\varphi)}}{\sqrt{\left(\cos(\varphi) + t \cdot \cos\left(\frac{\varphi}{2}\right)\cos(\varphi)\right)^2 + \left(\sin(\varphi) + t \cdot \cos\left(\frac{\varphi}{2}\right)\sin(\varphi)\right)^2 + \left(z - t \cdot \sin\left(\frac{\varphi}{2}\right)\right)^2}} \, dt \, d\varphi
$$

A numerical calcuation and comparison with the potential of a spherically symmetric charge (total charge equal) sitting in the origin, looks as shown in figure (3.1).

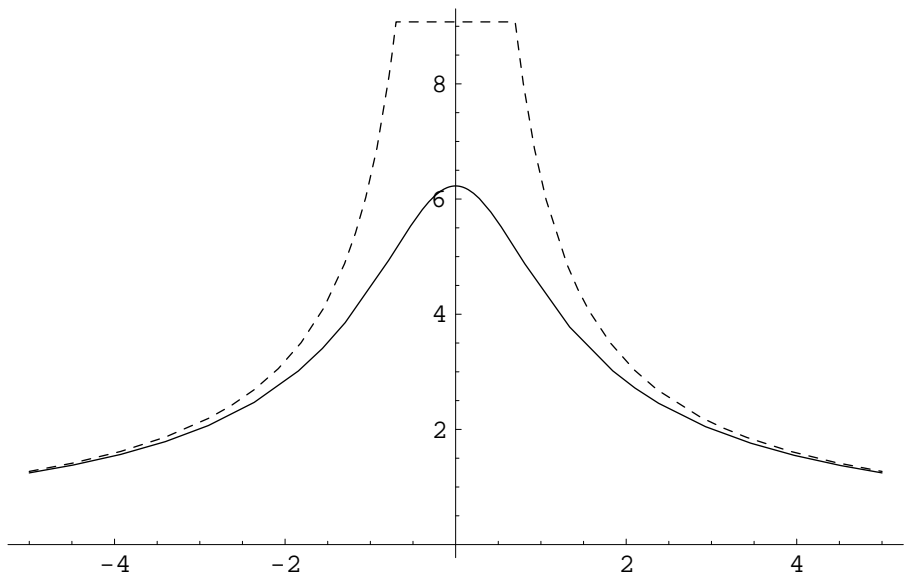

Figure 3.1: Potential on the z–axis of the Möbius strip (continuous line) and a spherically symmetric charge with the same total charge (dashed line)

### 3.2.2   Potential in 3D

To visualize the electrostatic Potential of a two–dimensional structure the *Mathematica*–package `AreaPotential.m` was programmed.  The *Mathematica* commands of the following explanation can be found in the file `session1.nb`.

As usual the package is loaded typing

*In[2]:=* **<< AreaPotential.m .**

To get a short description of what is included in this package, one just has to type

*In[3]:=* **?AreaPotential**

```
 contains :
  SetArea
  SetDistribution
  SetPotentialLaw
  GridValues
  Potential
  ViewPotential
  AnimatePotential
```

and you get a list of the functions provided by `AreaPotential.m`. For details of any of the functions listed above, again one enters a question mark and the name of the function.

Using the function `SetArea` one can define the area of interest for the package, i.e. its parametrisation. For convenience in programming the two parameters describing the surface are fixed to be *a* and *b*. Their range has also to be specified in the function `SetArea`. This rectangular area in parameter–space has to be split into small sub–rectangles doing the numerical calculation. The number of pieces per length can also be set by the user, it's the last argument of this function (its square is the total number of sub–rectangles). The default–value of this optional argument is 30.

A possible command to set the Möbius strip is therefore

```
In[4]:= SetArea[
            {Cos[b] + a Cos[b/2] Cos[b], Sin[b] + a Cos[b/2] Sin[b], a Sin[b/2]},
            {a, -0.5, 0.5}, {b, 0., 2. π}, 40] .
```

`SetDistribution` provides the possibility to set the area–charge–density, in dependence of the two area–parameters *a* and *b*. If one chooses a constant density it would be useful to set it in such a way, that the total charge is equal to 1. In the case of the Möbius strip this would be

$$In[5]:= \textbf{SetDistribution}\Big[\frac{\textbf{1}}{\textbf{6.35327}}\Big]$$

(see eq. 3.3). For generality the package allows the user to specify the potential law, e.g. $1/r$ for electrostatics (which is default) or $-\frac{g_0}{r}e^{-\mu r}$ for strong interaction. Doing this by using `SetPotentialLaw`, it is important to construct a formulation in cartesian coordinates with variables *x*, *y* and *z*. To clearify this the input for the $1/r$–law is given, although it's not necessary to enter it, for it's the default law:

$$In[6]:= \textbf{SetPotentialLaw}\Big[\frac{\textbf{1.}}{\sqrt{\textbf{x}^2 + \textbf{y}^2 + \textbf{z}^2}}\Big] .$$

Now nearly everything necessary is specified, the only remaining information for a numeric calculation is, in which point one wants to know the potential. If the aim is to visualize the potential graphically, one has to know the potential on a more or less dense grid of points. For this purpose the function `GridValues` can be used.

To specify the points on which the potential should be calculated, `GridValues` expects ranges for *x*, *y* and *z* and the number of intervals into which each of the ranges will be split to construct a grid of points. The last necessary argument is the maximum amount of the potential which is needed because of numerical reasons (see appendix B.1).

*In[7]:=* **Grid = GridValues[{-1.8,1.8},{-1.8,1.8},{-1.8,1.8},29,1.2];**

The back–value is a list of lists of lists of potential values. With this one can get
the potential on the point

$$\left( \begin{array}{c} x \\ y \\ z \end{array} \right) = \left( \begin{array}{c} -1,8 + (i-1) \cdot 0,0965517 \\ -1,8 + (j-1) \cdot 0,0965517 \\ -1,8 + (k-1) \cdot 0,0965517 \end{array} \right)$$

(see the arguments of `GridValues`) typing

*In[8]:=* **Grid[[i,j,k]] .**

Instead of working with this grid directly the package `AreaPotential.m` pro-
vides some more convenient functions.

For getting the potential on a point $(x, y, z)^{\mathrm{T}}$ use the function `Potential`, just
give the coordinate–vector as argument. The function will only return a value, if
the point lies within the ashlar specified when calling `GridValues`. The advan-
tage of this function compared to the list of grid–values is, that it automatically
interpolates for points other than grid points (see appendix B.1).

For the visualisation of the potential the function `ViewPotential` can be used.
The specified ashlar was devided into small intervals which means that the ash-
lar is also sliced into planes. For these planes the potential can be showed in a
3D–graphic. `ViewPotential` does so, expecting the index of the plane as an
argument. It has also the option `plane` which can be set to one of the values
`xy`, `xz` or `yz` to specify to which coordinate–axes the plane of interest shall be
parallel.

The great advantage of this package–function is the additional graphic which ap-
pears to the right. It shows the charged area and the plane for which the potential is
shown in the main picture. For the Möbius strip see the outputs of the commands

*In[9]:=* **ViewPotential[14,Grid,plane → xy] ,**

*In[10]:=* **ViewPotential[15,Grid,plane → xz]**

and

*In[11]:=* **ViewPotential[15,Grid,plane → yz]**

shown in fig. 3.2 to 3.4.

Of course the graphical description of the potential can, if one isn't interested in
the corresponding position of the "viewing plane", be easily produced by

*In[12]:=* **Plot3D[Potential[{x,y,0.}],{x,-1.8,1.8},{y,-1.8,1.8}] ,**

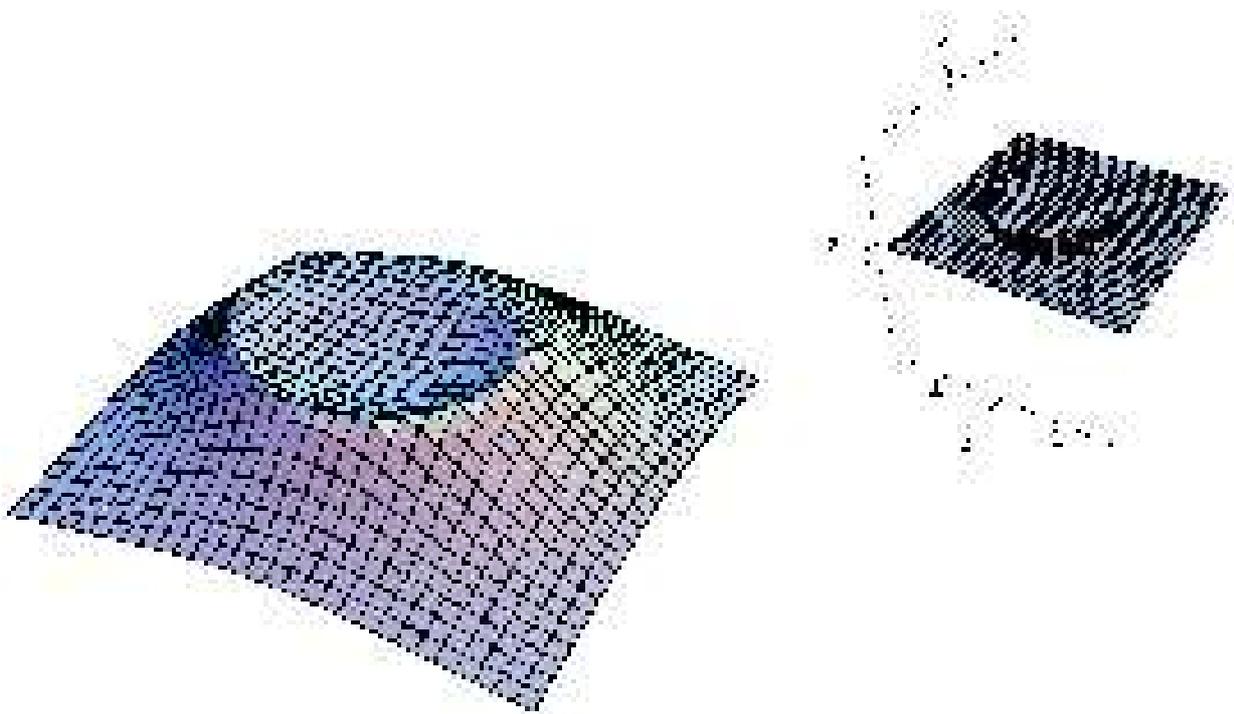

Figure 3.2: xy–graphic

its output can be seen in fig. 3.5.

There is another function provided by `AreaPotential.m` which visualises the potential by animating a sequence of potential surfaces and their corresponding "viewing planes". Naturally this cannot be shown in a printed form and its description is therefore left out, see the notebook `session1.nb` for its usage.

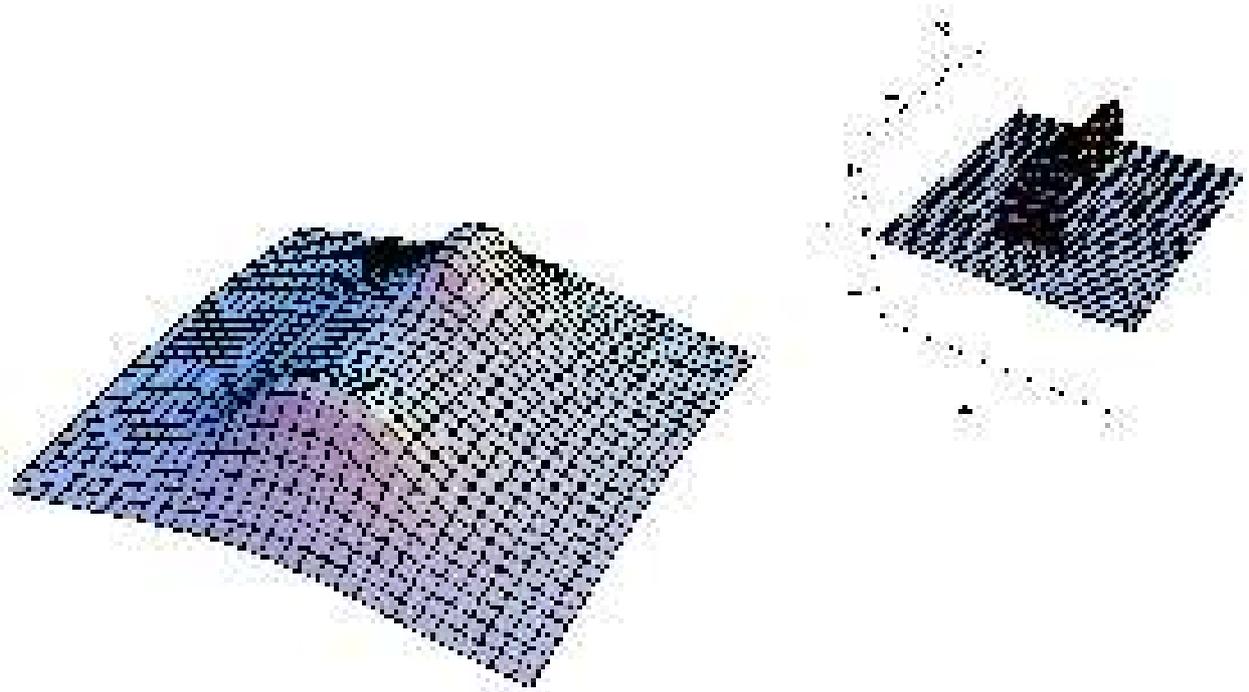

Figure 3.3: xz–graphic

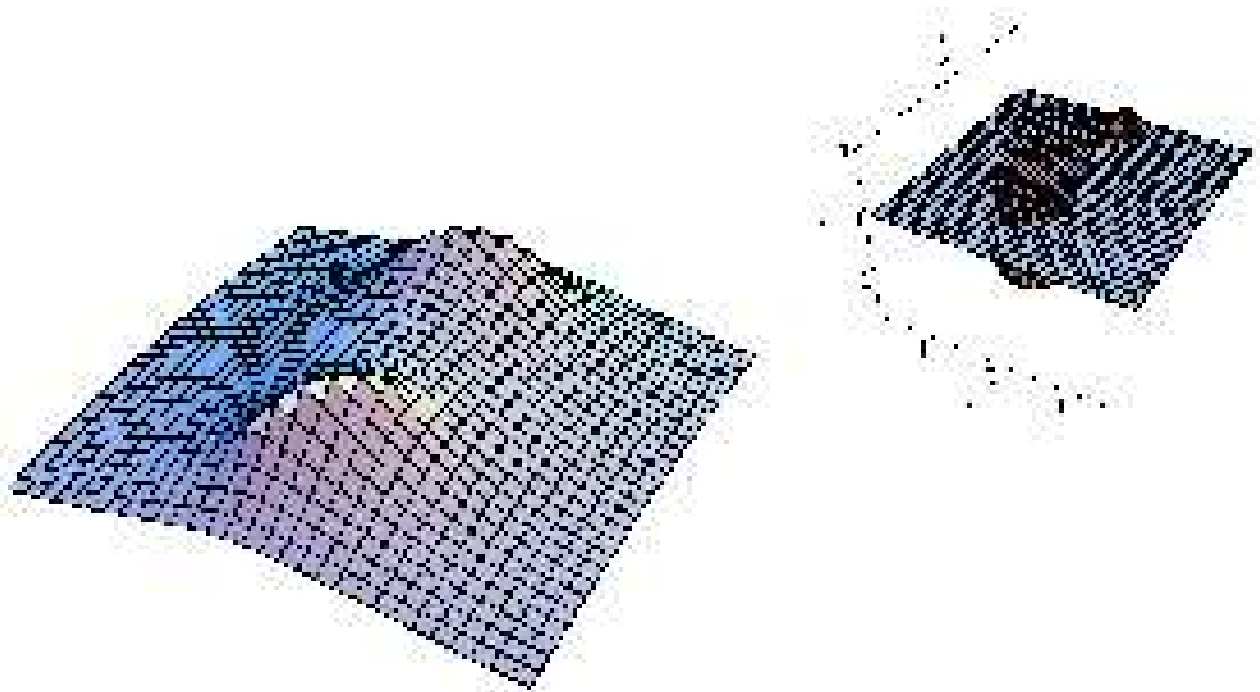

Figure 3.4: yz–graphic

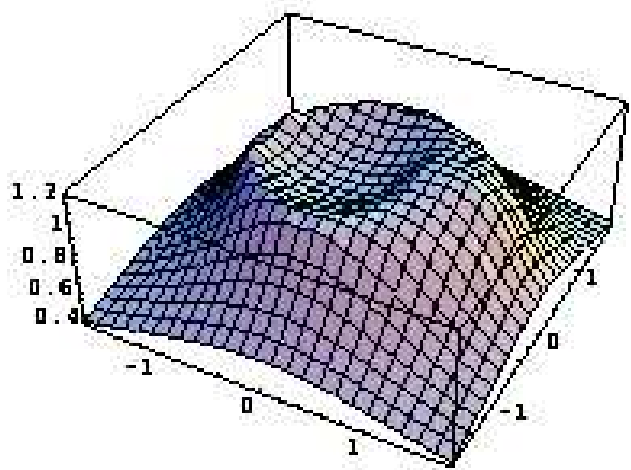

Figure 3.5: Potential of the Möbius strip

# Chapter 4

# Metallic Surfaces

If metallic surfaces are present, a charge put on this surface will distribute itself in a way that the total field–energy becomes a minimum. Equivalently one can say, the net–force on a single electron cannot have a component lying in the surface (i.e. its tangent plane) if the equilibrium is reached, which again is equivalent to "all positions are kept". This is also guaranteed by demanding the electric field (force per unit charge) always to be orthogonal to the surface (see [5, 2]).

It is the aim of this chapter to explain how the *Mathematica*–package `AreaConductor.m` can be used to find and display this configuration for a given surface and total charge.

Again the package can be loaded into the kernel by entering

*In[1]:=* **<< AreaConductor.m**

and a list of the contained functions is accesible as usual through

*In[2]:=* **?AreaConductor**

```
contains :
 SetArea
 SetPotentialLaw
 ChargePositions
 Potential
 SetCharges .
```

The first two commands are equivalent to those of `AreaPotential.m` (see 3.2.2 and B.1), the only difference in `SetPotentialLaw` being that now the variable has to be `r` instead of the triple (`x,y,z`). `ChargePositions` is the main command of this package, it does the charge–distribution–calculation numerically. To achieve this as fast as possible the numerical calculations are again exported from *Mathematica* to a plain C–program (see appendices B.2, C.2 and
D.2). The main idea of how to calculate the equilibrium shall be given here, al-
though it was implemented in the C–code and is not necessary for understadig
the usage of `AreaConductor.m`.

# 4.1   Finding the Equilibrium

It is of course possible to simulate the movement of a number of point charges
on a conducting surface by calculating the force and through this the acceleration
of each particle at any time step. One can then give the starting positions and
velocities and see how the system evolves.

The problem by doing so is, the particles will only stay in the equilibrium config-
uration if their momentary velocity is 0, otherwise they move on and start a kind
of oscillation. Even if one implements some damping or friction it will take a lot
of calculation–time–steps to reach the desired equilibrium.

Therefore a simulation of the "real" movement of the point charges is not the
proper way if one is only interested in the static equilibrium configuration. A
quite simple possibility to achieve the goal is to stop the particles after each time–
step.

It goes like this:

1. From one configuration (3D–positions of the particles) calculate the force
   on each of the particles.

2. Move each particle as if it were under influence of a constant force (the
   previously calculated), which means a distance proportional to the absolute
   value of the force ($\frac{1}{2}t^2/m$ is the proportionality factor which can numerically
   be set to some convenient value) into the direction of the momentary force–
   vector.

   For this movement, make sure the particle doesn't leave the surface doing
   two things:

   (a) Take only the force–component lying in the tangent plane of the sur-
       face and perform the shift of positions in parameter space, which means
       the particle is moved *along the surface* the determined distance.

   (b) If the particle moves outside the rectangular parameter range move it
       backwards to the edge it passed. (The method doesn't consider edges
       glued together, which is no problem because the particles can still

reach any place on the surface, it only probably takes a little longer to get there.)

3. Knowing the new parameter–values use the parametrisation to get the corresponding 3D–positions.

4. Start from beginning.

This procedure was programmed in C, the code can be seen in appendix D.2.

## 4.2 Numerical Approach

As promised the Möbius strip will be used for demonstration. The function `ChargePositions` can be used in two ways, distinguished by the given arguments. Two arguments are always mandatory, either one has to specify the number of point particles and the number of steps (see explanation above), or the arguments have to be a list of (guessed) positions in *parameter–space* and again the number of steps.

In either usages one can add the optional argument of a numerical (real) factor, which is a measure for the factor used to calculate the distance a particle is moved in one step (see above).

It is a good idea to use the possibility of choosing the force–distance–factor, one can then move the particles relatively "fast" from the initial positions (which are usually far away from equilibrium) and then gradually "slow down" the movement to prevent "jumping around" the equilibrium.

Using *Mathematica*'s clearity one can make a list of the number of steps and the corresponding movement–factors

```
In[3]:= parameters = {{3,1.},{10,1.},{10,0.9},{10,0.8},{10,0.7},
        {10,0.5},{10,0.1},{10,0.05},{10,0.03},{10,0.02},{10,0.01}} .
```

The first element with the 3 steps is only for a better visualisation of the movement and not useful if one is only interested in the final result.

The return–value of `ChargePositions` is a list of (first) the particle–positions in 3D–space (cartesian coordinates) and (second) the particle–positions in 2D–parameter–space. For a recursive application of `ChargePositions` one has to define the following function:

```
In[4]:= f[L_,{n_,q_}] := ChargePositions[L[[2]],n,q] .
```

Now one can use `FoldList` to produce a list of charge positions resulting from a successive application of `f` on the last result, taking successive elements of `parameters` as parameters. The initial positions are left to the program, it is told to distribute 2500 particles over the surface.

*In[5]:=* **l = FoldList[f, ChargePositions[2500, 0], parameters];**

The result is now best visualised by

*In[6]:=* **AnimateGraphics[(Show[Graphics3D[Point /@ #[[1]]]])& /@ l]**

which, in this case, produces an animation based on the graphics seen in fig. 4.1.

It is often useful to save the results, so that one doesn't have to re calculate a quite time consuming distribution. In this case the ".dat" extension of the *Mathematica* command `Export` is quite easy to handle:

*In[7]:=* **Export["pos2500 – 3D.dat", Last[l][[1]]] ,**

*In[8]:=* **Export["pos2500 – 2D.dat", Last[l][[2]]] .**

The saved data can be retrieved for instance by

*In[9]:=* **l2D = Import["pos2500 – 2D.dat"]**

or

*In[10]:=* **l3D = Import["pos2500 – 3D.dat"] .**

Any imported 3D–list of data can be loaded into the Potential–function (the last output of `ChargePositions` is automatically loaded):

*In[11]:=* **SetCharges[l3D] .**

One can then get the potential just typing e.g.

*In[12]:=* **Potential[{5, 2, 3}]**
*Out[12]=* **0.162799 .**

There is, as one can see, no restriction on the point–vector, for the potential of a sum of point charges is of course extrapolatable to an arbitrary position. To compare the potential of the charged metallic Möbius strip with the homogeneously charged one we visualise the potential in a similar way:

*In[13]:=* **Plot3D[Potential[{x, y, 0.}], {x, -1.8, 1.8}, {y, -1.8, 1.8}] ,**

the output can be seen in fig. 4.2.

A comparison with fig. 3.5 shows that the maximum value of the potential of the metallic strip is slightly smaller in the chosen plane ($z = 0$), because the charges are driven to the rim of the strip. This is also the reason why the potential in the section where the strip is parallel to the x–y–plane is flatter (the metallic strip is an equipotential–area).

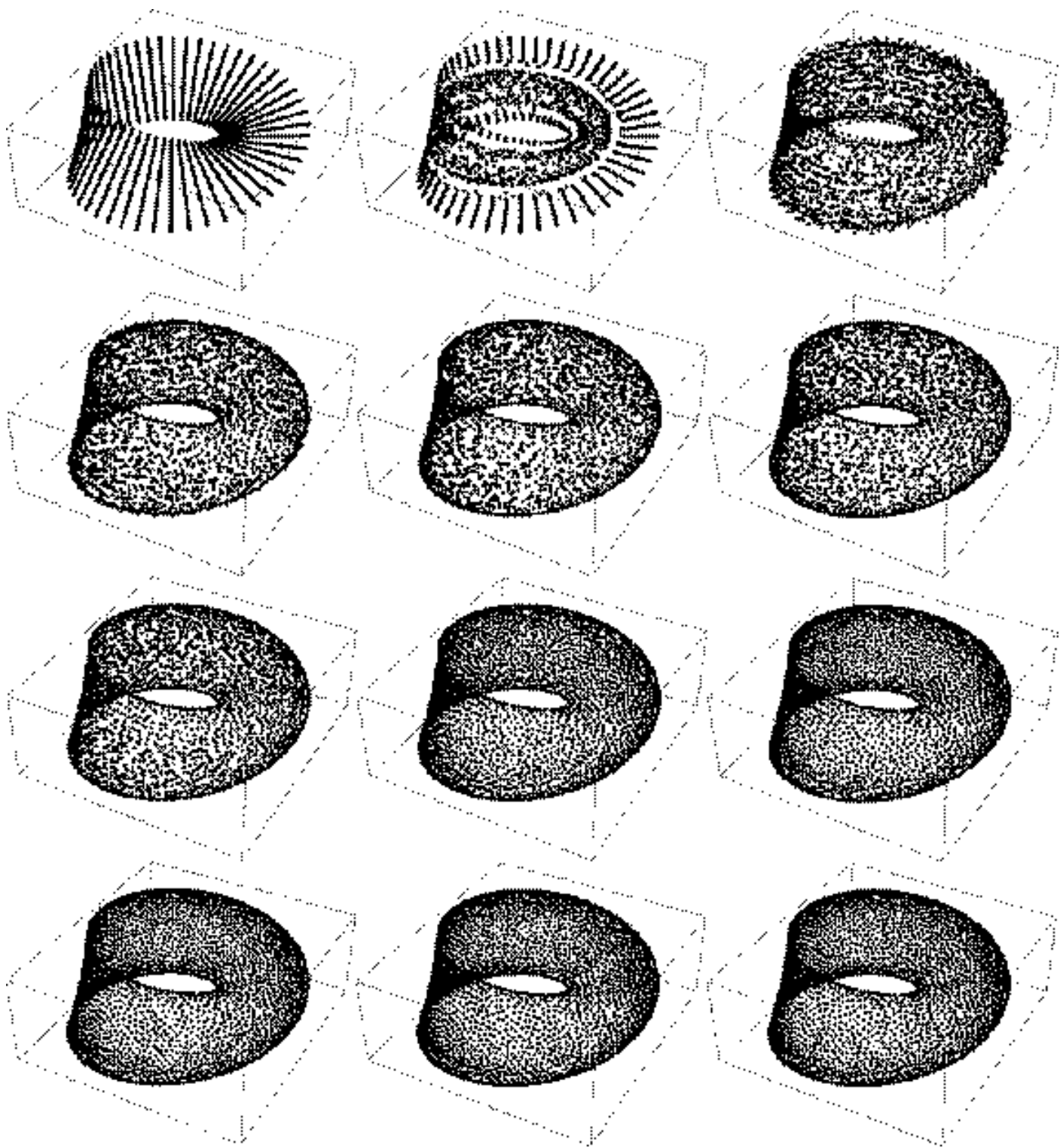

Figure 4.1: Distributions of point–particles on the Möbius strip .

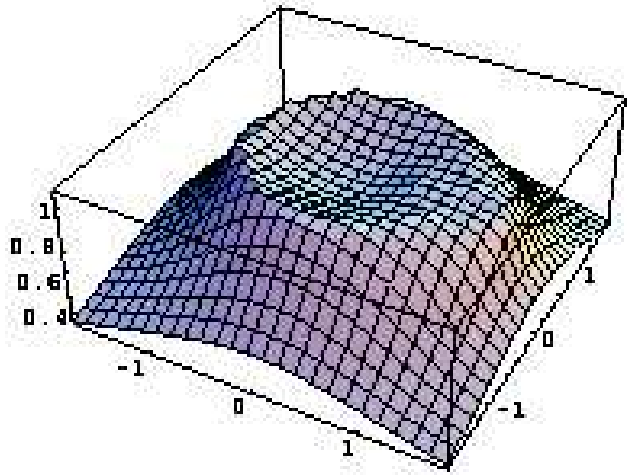

Figure 4.2: Potential of a metallic, charged Möbius strip

# Chapter 5

# Radiation

## 5.1 Fundamentals

The following considerations can be learned in more detail in e.g. [2] or [5].

For localised charges and currents the vectorpotential $A_i$ can be calculated with the appropriate Green's function:

$$A_i\left(r_j, t\right) = \frac{1}{c} \int d^3 r' \frac{j_i\left(r'_j, t - \frac{|r_j - r'_j|}{c}\right)}{|r_j - r'_j|} \ .$$

Being only interested in the fields far away from their sources ($r' \ll r$) the magnetic field can (using $B_i = \varepsilon_{ijk}\partial_j A_k$) be written as

$$B_i = -\varepsilon_{ijk}\frac{r_j}{c^2 r^2} \int d^3 r' \dot{j}_k\left(r'_l, t - \frac{r}{c} + \frac{r_l r'_l}{cr}\right) + O\left(\frac{1}{r^2}\right)$$

where the short notation $\dot{j} = \frac{\partial j}{\partial t}$ is used. If one also introduces

$$\dot{q}_i\left(r_j, t\right) = \int d^3 r' \ j_i\left(r'_j, t - \frac{r}{c} + \frac{r_j r'_j}{cr}\right) \ , \tag{5.1}$$

the so called radiation part of the magnetic field is

$$B_{ri}\left(r_j, t\right) = \frac{\varepsilon_{ijk}\ddot{q}_j r_k}{c^2 r^2} \tag{5.2}$$

and, using the Maxwell–equation in the (source free) radiation region

$$\frac{1}{c}\frac{\partial E_j}{\partial t} = \varepsilon_{ijk}\partial_j B_k$$

the radiation part of the electric field can be expressed as

$$E_{\text{r}\,i}\left(r_j, t\right) = \frac{\varepsilon_{ijk}(\varepsilon_{jlm}\dddot{q}_l r_m)r_k}{c^2 r^3}\ . \tag{5.3}$$

Now the radiation part of the Poynting vector can be defined:

$$S_{\text{r}\,i}\left(r_j, t\right) = \frac{c}{4\pi}\varepsilon_{ijk}E_{\text{r}\,j}\left(r_j, t\right)B_{\text{r}\,k}\left(r_j, t\right)\ .$$

Inserting eq. 5.2 and eq. 5.3 one gets ($e_i^r$ is the $i^{th}$ cartesian component of the unit vector in radial direction)

$$S_{\text{r}\,i}\left(r_j, t\right) = \frac{c}{4\pi}\frac{1}{c^4 r^2}\left(\dddot{q}^2 - \left(e_j^r\dddot{q}_j\right)^2\right)e_i^r \tag{5.4}$$

or, in vector notation

$$\mathbf{S}_{\text{r}}\left(\mathbf{r}, t\right) = \frac{c}{4\pi}\frac{1}{c^4 r^2}\left(\dddot{\mathbf{q}}\times \mathbf{e}_r\right)^2\mathbf{e}_r\ . \tag{5.5}$$

The quantity of interest is the radiation power with respect to the (radial) direction, i.e. the radiation power per infinitesimal area–element of the unit sphere. Using $dP = \mathbf{S}\cdot d^2\mathbf{f}$ one easily gets

$$\frac{dP}{d\Omega} = \frac{1}{4\pi c^3}\left(\dddot{\mathbf{q}}\times \mathbf{e}_r\right)^2 \tag{5.6}$$

which is a quite easy formula if one only knows $\mathbf{q}$. The standard way for calculating $\mathbf{q}$ uses the following considerations.

The function $\mathbf{q}\left(\mathbf{r}, t\right)$ can also be seen as a function of the direction $\mathbf{e}_r = \frac{\mathbf{r}}{r}$ and the retarded time $t_r = t - \frac{r}{c}$, it's written as $\mathbf{q}\left(t_r, \mathbf{e}_r\right)$ (mark the exchange of order of "time" and "space" arguments). In the resulting integral

$$\dot{\mathbf{q}}\left(t_r, \mathbf{e}_r\right) = \int d^3 r'\ \mathbf{j}\left(\mathbf{r}', t_r + \frac{\mathbf{r}'\cdot \mathbf{e}_r}{c}\right)$$

one takes the first two elements of the Taylor expansion around $t_r$ (zero and first order) of the integrand $\mathbf{j}$:

$$\dot{\mathbf{q}}(t_r, \mathbf{e}_r) = \int d^3 r' \, \mathbf{j}(\mathbf{r}', t_r) + \int d^3 r' \, \dot{\mathbf{j}}(\mathbf{r}', t_r) \frac{\mathbf{r}' \cdot \mathbf{e}_r}{c} \, .$$

Using $j_i = \partial_j j_j r_i - (\partial_j j_j) r_i$, the continuity equation $\dot{\rho} + \partial_i j_i = 0$ and the fact that the sources are localised, one can simplify the first integrand to $\mathbf{r}' \dot{\rho}(\mathbf{r}', t_r)$.

A little more tricky is the transformation of the second integrand. Firstly, similar to the transformation of the first integrand one wants a total derivative, which will vanish due to locality. The corresponding equality would be

$$\partial_i' j_i r_j r_j' r_k' = \underbrace{(\partial_i' j_i)}_{-\dot{\rho}} r_j r_j' r_k' + j_i r_i r_k' + j_k r_j r_j' \, . \tag{5.7}$$

Now one realizes that the last two terms are also obtained by

$$\varepsilon_{kji} \varepsilon_{jlm} r_l' j_m r_i = r_i' j_k r_i - r_k' j_i r_i \, . \tag{5.8}$$

Adding eq. 5.7, eq. 5.8 and $\dot{\rho} \, r_j r_j' r_k'$ results in the desired integrand, up to some factors. Therefore an equivalent formula for the $\mathbf{q}$–vector is

$$\begin{aligned}
\mathbf{q}(t_r, \mathbf{e}_r) &= \int d^3 r' \, \mathbf{r}' \rho(\mathbf{r}', t_r) + \frac{1}{2c} \int d^3 r' \, (\mathbf{r}' \times \mathbf{j}(\mathbf{r}', t_r)) \times \frac{\mathbf{r}}{r} + \\
&\quad + \frac{1}{2cr} \int d^3 r' \, \dot{\rho}(\mathbf{r}', t_r)(\mathbf{r} \cdot \mathbf{r}') \mathbf{r}' \, .
\end{aligned} \tag{5.9}$$

With the definitions

$$\mathbf{p}(t_r) = \int d^3 r' \, \mathbf{r}' \rho(\mathbf{r}', t_r) \, , \tag{5.10}$$

$$\mathbf{m}(t_r) = \frac{1}{2c} \int d^3 r' \, \mathbf{r}' \times \mathbf{j}(\mathbf{r}', t_r) \quad \text{and} \tag{5.11}$$

$$Q_{ij}(t_r) = 3 \int d^3 r' \, r_i' r_j' \rho(\mathbf{r}', t_r) \tag{5.12}$$

of the electric dipole momentum (5.10), the magnetic dipole momentum (5.11) and the electric quadrupole momentum (5.12), $\ddot{\mathbf{q}}$ can, in this approximation, be expressed as

$$\dddot{\mathbf{q}}\left(t_r, \mathbf{e}_r\right) = \dddot{\mathbf{p}}(t_r) + \dddot{\mathbf{m}}(t_r) \times \mathbf{e}_r + \frac{1}{6c}\mathbf{e}_r^{\mathsf{T}}\,\dddot{\mathbf{Q}}(t_r)\,. \qquad (5.13)$$

For the validity of this approximation the charge density (or the current density) must not vary significantly in the time $\frac{\mathbf{r}\cdot\mathbf{r}'}{cr}$. This term is of order $\frac{d}{c}$, where $d$ is the characteristic diameter of the charge and current distribution. If the characteristic time for a change of $\rho$ or $\mathbf{j}$ is $T$, the necessary condition is

$$\frac{d}{c} \ll T\,.$$

Another way for formulating this condition is to set the magnitude of the velocity of the moved charges to $v$, which can be compared with $\frac{d}{T}$ and therefore one gets

$$v \ll c\,.$$

This shows that the approximation describes a **nonrelativistic** movement.

## 5.2   Application to Rotation

To guarantee the validity of eq. 5.10 to eq. 5.13 for a rotating system one has to fulfil

$$d \ll \lambda$$

or

$$f \ll \frac{c}{d}\,,$$

$\lambda \approx cT$ being the characteristic wavelength and $f = 1/T$ the characteristic frequency.

### 5.2.1   Rotation of an Ensemble of Point Charges

To use the outcome of section 4.2 for further calculations some ideas about treating rotations are given now, influenced by [7] (see also [1]).

A rotating vector $\mathbf{r}(t)$ can in general be described as

$$\mathbf{r}(t) = \mathbf{B}(t)\mathbf{r}_0 \, ,$$

$\mathbf{r}_0$ being the initial position and $\mathbf{B}(t)$ the transformation matrix in dependency of time, the rotation matrix ($\mathbf{B}(0) = \mathbb{1}$). The well kown matrices for rotation around the $z$– and $y$–axis will be needed. With $\varphi$ being the rotation angle they are

```
In[1]:= Bz[φ_] := ⎛ Cos[φ]  -Sin[φ]  0 ⎞
                  ⎜ Sin[φ]   Cos[φ]  0 ⎟
                  ⎝   0        0      1 ⎠
```

and

```
In[2]:= By[φ_] := ⎛  Cos[φ]  0  Sin[φ] ⎞
                  ⎜    0      1    0    ⎟ .
                  ⎝ -Sin[φ]  0  Cos[φ] ⎠
```

The most general problem for an ensemble of point charges would be a list of charge positions and an arbitrary rotation axis with angular velocity $\omega$.

Usually the axis will include the origin of the coordinate system. If not this can easily be made so by subtracting a vector from the origin to an in general arbitrary point of the axis (the new origin) from all position vectors of the point charges.

Now the rotation axis still has an arbitrary direction, described by the angles $\theta_\omega$ and $\varphi_\omega$. For an easy description of radiation it would be most convenient to have e.g. the $z$–axis as rotation axis. This can be achieved by using a new coordinate system which exactly fulfils this. To get the initial positions of the charges in this coordinate system one has to multiply each position–vector by a simple transformation matrix given by

```
In[3]:= B[θomega_, φomega_] := By[-θomega].Bz[-φomega] .
```

To take the conducting Möbius strip as an example one can use the saved data (see 4.2) and e.g. the $z$–axis of the standard parametrisation as rotation axis, the positions are then

```
In[4]:= positions = (B[0,0].#)&/@Import["pos2500-3D.dat"] .
```

Before calculating the electro–magnetic momenta of interest one can make these calculations easier by some more considerations.

When using eq. 5.10 on point charges one gets

$$\mathbf{p}(t_r) = \int d^3r' \ \mathbf{r}' \sum_{n=1}^{N} q_n \delta\left(\mathbf{r}' - \mathbf{r}_n(t_r)\right) \, .$$

$N$ is the total number of point–particles, $\mathbf{r}_n(t_r)$ are their positions at time $t_r$. The $q_n$ are the charges that are from now on taken to be equal and for simplicity set to $1/N$, which sets the total charge to 1.

The positions $\mathbf{r}_n(t_r)$ can be expressed from the initial positions $\mathbf{r}_{n0}$ and the rotation matrix around $z$ with argument $\omega t_r$. Due to the linearity of the integral the sum can be pulled out which yields

$$
\begin{aligned}
\mathbf{p}(t_r) &= \frac{1}{N}\sum_{n=1}^{N}\int d^3r'\, \mathbf{r}'\delta\left(\mathbf{r}' - \mathbf{B}_z(\omega t_r)\mathbf{r}_{n0}\right) = \frac{1}{N}\sum_{n=1}^{N}\mathbf{B}_z(\omega t_r)\mathbf{r}_{n0} = \\
&= \mathbf{B}_z(\omega t_r)\frac{1}{N}\sum_{n=1}^{N}\mathbf{r}_{n0} = \mathbf{B}_z(\omega t_r)\mathbf{p}_0
\end{aligned}
$$

with $\mathbf{p}_0$ being the initial electric dipole momentum. This can now be calculated for the example positions:

```
In[5]:= pZero = 1 / Length[positions] Apply[Plus, positions]

Out[5]= {0.00530198, 0.00202618, 0.00111019} .
```

Of main interest is the *mean* radiation power per $d\Omega$, which means the average over one period $2\pi/\omega$ will be taken. As a consequence one can just type $t$ instead of $t_r$, the shift of $r/c$ being of no influence. The electric dipole momentum is therefore

```
In[6]:= p = Bz[ω t].pZero

Out[6]= {0.00530198 Cos[tω] - 0.00202618 Sin[tω] , 0.00202618 Cos[tω] +
0.00530198 Sin[tω] , 0.00111019} .
```

For later comparison the radiation of just this contribution will be visualised. To do so this vector has to be differentiated twice with respect to time, then the outer product with the vector $\mathbf{e}_r$ has to be calculated and the square of the result has to be taken. In the result the average over one period can be calculated by using

$$
\begin{aligned}
\left\langle \sin^2(\omega t)\right\rangle &= \left\langle \cos^2(\omega t)\right\rangle = \frac{1}{2}\,, \\
\left\langle \sin^4(\omega t)\right\rangle &= \left\langle \cos^4(\omega t)\right\rangle = \frac{3}{8}\,, \\
\left\langle \sin^2(\omega t)\cos^2(\omega t)\right\rangle &= \frac{1}{8}\,, \\
\left\langle \sin(\omega t)\cos^i(\omega t)\right\rangle &= \left\langle \sin^i(\omega t)\cos(\omega t)\right\rangle = 0 \quad \text{and} \\
\left\langle \sin^o(\omega t)\right\rangle &= \left\langle \cos^o(\omega t)\right\rangle = 0\,,
\end{aligned}
$$

where *i* is an arbitrary integer and *o* an arbitrary odd integer. In *Mathematica* code one has

```
In[7]:= meandPdΩ1[θ_,φ_,ω_,c_] :=
           Evaluate[
             Collect[
               Simplify[
                 Expand[ 1/(4 π c³) Apply[Plus, Cross[D[p, {t, 2}],
                                          {Sin[θ] Cos[φ], Sin[θ] Sin[φ],
                                           Cos[θ]}]^2]]/.
                   x_?NoT Sin[y_t]⁴ → x 3/8 /.
                   x_?NoT Sin[y_t]² → x 1/2 /.
                   x_?NoT Cos[y_t]⁴ → x 3/8 /. x_?NoT Cos[y_t]² →
                   x 1/2 /. x_?NoT Sin[y_t]² Cos[y_t]² → x 1/8 /.
                   x_?NoT Sin[y_t] Cos[y_t]^-Integer → 0 /.
                   x_?NoT Cos[y_t] Sin[y_t]^-Integer → 0 /.
                   x_?NoT Sin[y_t]^n_?OddQ → 0 /.
                   x_?NoT Cos[y_t]^n_?OddQ → 0 /. x_?NoT Sin[y_t] → 0 /.
                   x_?NoT Cos[y_t] → 0 /. x_ Sin[φ]² → x (1 - Cos[φ]²) /.
                   x_ Sin[θ]² → x (1 - Cos[θ]²)]/. 0. x_ → 0., c]] .
```

`NoT` is just a function which checks that its argument doesn't contain the symbol `t`:

```
In[8]:= NoT[x_] := FreeQ[x, t] .
```

To clearify this long function:

```
In[9]:= meandPdΩ1[θ, φ, ω, c]
```
$$Out[9]= \frac{1.28185 \times 10^{-6}\, \omega^4 + 1.28185 \times 10^{-6}\, \omega^4\, Cos[\theta]^2}{c^3} \,.$$

Now the following numerical values are taken: $c = 3 \cdot 10^{10}$ and $\omega = 3000$. Units are left out in this *Mathematica*–session (it would be cm/s and s$^{-1}$ for these quantities). To get easily readable and easily visualisable values the function `g` (for "graphical") is introduced:

```
In[10]:= g[θ_,φ_,ω_,c_] := 10²⁴ meandPdΩ1[θ, φ, ω, c] .
```

One gets for example

```
In[11]:= g[0, 0, 3000, 3 10¹⁰]
```

*Out[11]=* 7.69109 .

Now the radiation power dependent of direction can be visualised. A quite revealing method is to plot the points the following way: Take a direction $(\theta, \varphi)$ and multiply the unit vector in this direction with the corresponding radiated power.

The result is an area which contains exactly the desired information. If one takes a point on the area the radiated power in this direction (from the origin) is the distance from the origin.

*In[12]:=* `ParametricPlot3D[{g[`$\theta$`,`$\varphi$`, 3000, 3 10`$^{10}$`] Sin[`$\theta$`] Cos[`$\varphi$`],`
`g[`$\theta$`,`$\varphi$`, 3000, 3 10`$^{10}$`] Sin[`$\theta$`] Sin[`$\varphi$`],`
`g[`$\theta$`,`$\varphi$`, 3000, 3 10`$^{10}$`] Cos[`$\theta$`]}, {`$\theta$`, 0,`$\pi$`}, {`$\varphi$`, 0, 2`$\pi$`}]`

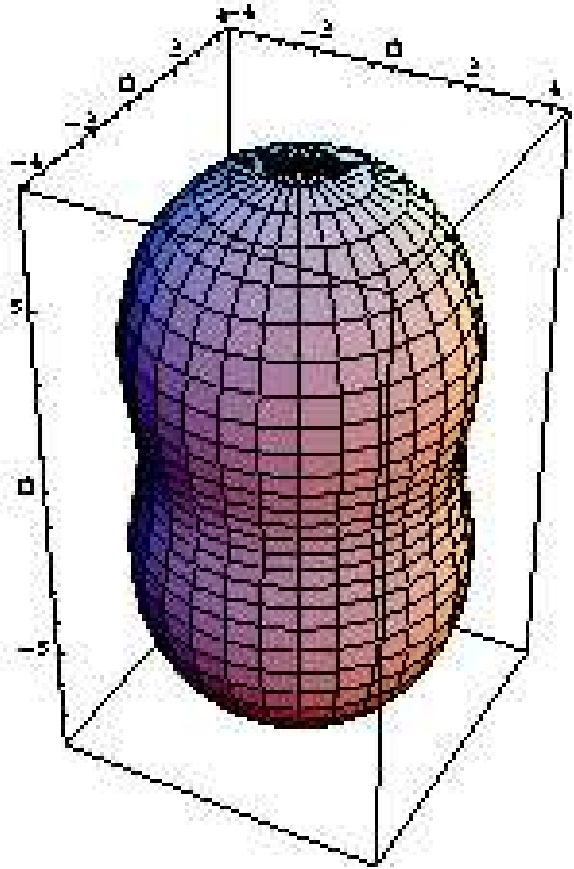

As expected the radiation shows no $\varphi$–dependence because $\varphi$ is the angle from the $x$–axis around the $z$–axis, which is the rotation axis. The average power is therefore independent of this angle. As a consequence it also makes sense only to do a 2D–plot in the $x$–$z$–plane ($\varphi = 0$):

```
In[13]:= ParametricPlot[{g[θ,0,3000,3 10¹⁰] Sin[θ],
            g[θ,0,3000,3 10¹⁰] Cos[θ]},{θ,0,2π},
          PlotRange→{{-8,8},{-8,8}},AspectRatio→Automatic]
```

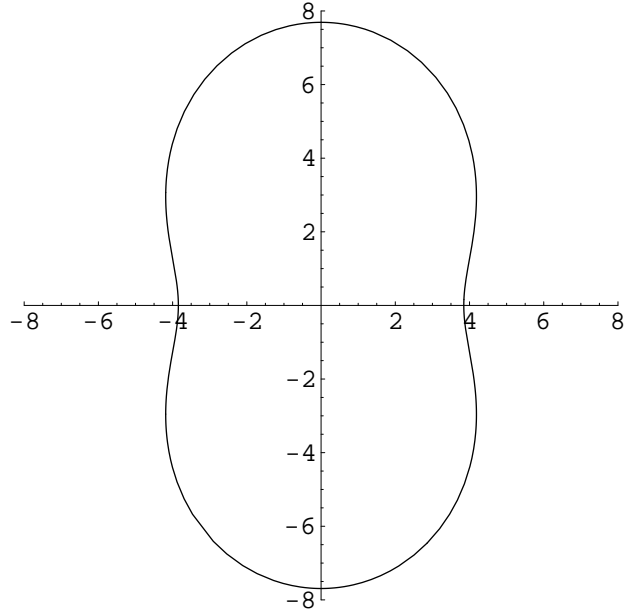

Now the magnetic dipole and electric quadrupole momenta shall be considered. The first steps are again simplifications of the integrals, for which $\mathbf{r}_n(t) = \mathbf{B}_z(\omega t)\mathbf{r}_{n0}$, $\mathbf{v}_n(t) = \dot{\mathbf{B}}_z(\omega t)\mathbf{r}_{n0}$ and (for each charge) $\mathbf{j} = \mathbf{v}\rho$ are needed. The following term occurs under the integral for $\mathbf{m}$ and can be transformed:

$$\mathbf{r} \times \mathbf{v} = (\mathbf{B}\mathbf{r}_0) \times (\dot{\mathbf{B}}\mathbf{r}_0) = \mathbf{B}\left(\mathbf{r}_0 \times (\mathbf{B}^{\mathrm{T}}\dot{\mathbf{B}}\mathbf{r}_0)\right).$$

Considering $(\mathbf{B}^{\mathrm{T}}\dot{\mathbf{B}}\mathbf{r}_0) = \boldsymbol{\Omega} \times \mathbf{r}_0$, $\boldsymbol{\Omega} = \mathbf{B}^{\mathrm{T}}\boldsymbol{\omega} = \boldsymbol{\omega}$ ($\boldsymbol{\omega}$ lies in the rotation axis) and $\boldsymbol{\omega} = \omega\mathbf{e}_z$ one can make the simplifications (index 0 for initial is dropped)

$$\varepsilon_{ijk}\varepsilon_{klm}r_j e_l^z r_m = \left(\delta_{il}\delta_{jm} - \delta_{im}\delta_{jl}\right)r_j e_l^z r_m = r_j e_i^z r_j - r_j e_j^z r_i = e_i^z \mathbf{r}^2 - r_i r_z\,.$$

$\mathbf{B}(t)$ rotates around the $z$–axis, therefore $\mathbf{B}(t)\mathbf{e}_z = \mathbf{e}_z$. As a consequence the first term of $\mathbf{m}(t)$ doesn't depend on time and is irrelevant for the (final) power calculation, where $\mathbf{m}(t)$ only appears two times differentiated w.r.t. time. For the sake of completeness it shall still be given here:

$$\mathbf{m}(t) = \frac{\omega}{2cN}\left(\mathbf{e}_z \sum_{n=1}^{N} \mathbf{r}_n^2 - \mathbf{B}_z(\omega t) \sum_{n=1}^{N} r_n^z \mathbf{r}_n\right),$$

```
In[14]:= m = ────────────ω────────────
            2 c Length[positions]
               ({0, 0, 1} Plus @@ (Norm /@ positions) -
                  Bz[ω t].(Plus @@ Map[(#[[3]] * #) &, positions]))
```

```
Out[14]= { ω (1.07295 Cos[t ω] + 205.004 Sin[t ω])
          ─────────────────────────────────────────── ,
                           5000 c
          ω (-205.004 Cos[t ω] + 1.07295 Sin[t ω])      0.530487 ω
          ──────────────────────────────────────── ,   ───────────── } .
                         5000 c                              c
```

For later calculations the symbol $\mathbf{m}'$ is introduced which represents mainly the second term of $\mathbf{m}$, but leaves out the rotation matrix and includes on the other side the prefactor $1/N$, which makes it a simple numeric quantity:

```
In[15]:= mPrime = ──────1──────── (Plus@@Map[(#[[3]]*#)&, positions])
                  Length[positions]
```

```
Out[15]= {-0.000429178, 0.0820015, 0.0665318} .
```

The electric quadrupole can similarly be simplified:

$$Q_{ij}(t) = \frac{3}{N}\sum_{n=1}^{N} r_{ni}(t)r_{nj}(t) = \mathbf{B}_{zik}(\omega t)\underbrace{\left(\frac{3}{N}\sum_{n=1}^{N} r_{nk}r_{nl}\right)}_{=Q_{0kl}}\mathbf{B}_{zlj}^{\mathrm{T}}(\omega t),$$

```
In[16]:= Qzero = ──────3────────
                 Length[positions]
                   Apply[Plus, Map[(Outer[Times, #, #])&, positions]]
```

```
Out[16]= {{1.84205, -0.00466737, -0.00128754},
          {-0.00466737, 1.91751, 0.246004}, {-0.00128754, 0.246004, 0.199595}} .
```

With the definitions

```
In[17]:= Bdot = D[Bz[ω t], t];
         Bdotdot = D[Bdot, t];
         Bdotdotdot = D[Bdotdot, t];
```

and the introduction of a short form for transposition

```
In[18]:= T[q_] := Transpose[q]
```

$\overset{..}{\mathbf{q}}$ can be calculated:

```
In[19]:= qDoubleDot =
            Bdotdot.pZero-
              ω
             ─── Cross[Bdotdot.mStrich, {Sin[θ] Cos[φ], Sin[θ] Sin[φ], Cos[θ]}]+
             2 c
              1
             ─── {Sin[θ] Cos[φ], Sin[θ] Sin[φ], Cos[θ]}.
             6 c
                (Bdotdot.Q.Bz[-ω t] + 3 Bdotdot.Q.T[Bdot] + 3 Bdot.Q.T[Bdotdot]+
                    Bz[ω t].Q.T[Bdotdotdot]) .
```

And from that we can build

```
In[20]:= meandPdΩ2[θ_, φ_, ω_, c_] :=
            Evaluate[
              Collect[
                Simplify[
                  Expand[ ───────── (Plus @@ (Cross[qDoubleDot, {Sin[θ] Cos[φ],)
                           4 π c³
                                                Sin[θ] Sin[φ], Cos[θ]}]²) ] ]/.

                                 x_?NoT Sin[ω_ t]⁴ → x 3/8 /.

                                 x_?NoT Sin[ω_ t]² → x 1/2 /.

                                 x_?NoT Cos[ω_ t]⁴ → x 3/8 /.

                                 x_?NoT Cos[ω_ t]² → x 1/2 /.

                                 x_?NoT Sin[ω_ t]² Cos[ω_ t]² → x 1/8 /.

                                x_?NoT Sin[ω_ t] Cos[ω_ t]⁻ᴵⁿᵗᵉᵍᵉʳ → 0 /.
                                x_?NoT Cos[ω_ t] Sin[ω_ t]⁻ᴵⁿᵗᵉᵍᵉʳ → 0 /.
                              x_?NoT Sin[ω_ t]ⁿ_?ᴼᵈᵈQ → 0 /.
                             x_?NoT Cos[ω_ t]ⁿ_?ᴼᵈᵈQ → 0 /. x_?NoT Sin[ω_ t] → 0 /.
                           x_?NoT Cos[ω_ t] → 0 /. Sin[φ]² → (1 - Cos[φ]²) /.
                        Sin[θ]² → (1 - Cos[θ]²) /.
                       Cos[φ] Sin[φ]³ → Cos[φ] Sin[φ] (1 - Cos[φ]²) /.
                      Sin[φ]⁴ → (1 - Cos[φ]²)², 1/c ] //. z_ (0. x_ + y_) → y z //.
                   (0. + 0. i) x_ + y_ → y] .
```

This function is quite illustrating, because it is ordered in powers of $1/c$:

```
In[21]:= meandPdΩ2[θ, φ, ω, c]
```

$Out[21]=$ $\dfrac{1.28185 \times 10^{-6}\,\omega^4 + 1.28185 \times 10^{-6}\,\omega^4\, Cos[\Theta]^2}{c^3} +$

$\qquad\quad \dfrac{-0.0000346671\,\omega^5\, Cos[\Theta] - 0.0000346671\,\omega^5\, Cos[\Theta]^3}{c^4} +$

$\qquad\quad \dfrac{1}{c^5}\,(0.000523188\,\omega^6\, Cos[\Theta]^2 + 0.0000119254\,\omega^6\, Cos[\Theta]^4 +$

$\qquad\qquad 0.0000534376\,\omega^6\, Sin[\Theta]^4 + 0.000595024\,\omega^6\, Cos[\varphi]^2\, Sin[\Theta]^4 -$

$\qquad\qquad 0.000595024\,\omega^6\, Cos[\varphi]^4\, Sin[\Theta]^4 + 0.000074745\,\omega^6\, Cos[\varphi]\, Sin[\Theta]^4\, Sin[\varphi] -$

$\qquad\qquad 0.00014949\,\omega^6\, Cos[\varphi]^3\, Sin[\Theta]^4\, Sin[\varphi])\,.$

The terms of order $1/c^3$ are exactly the same as if one had simply a rotating dipole, as one should have expected. The $1/c^4$–terms are "mixed" terms of the electric dipole on one side and the magnetic dipole or the electric quadrupole on the other side, due to the square in the formula of $dP/d\Omega$. Terms of order $1/c^5$ originate qualitatively from "$\mathbf{m}^2$", "$\mathbf{Q}^2$" and "$\mathbf{m} \cdot \mathbf{Q}$".

It is clear that, as long as there is a dipole–contribution and the approximation $\omega \ll c$ (which is the nonrelativistic–condition in the case $d \approx 1$) holds, only the dipole is of interest. All terms of higher order in $1/c$ (i.e. $\omega/c$) are negligible.

This fact can also be visualised. (One again has to use an "inteface–function".)

$In[22]:=$ $\mathbf{f[\theta\_,\varphi\_] = Evaluate\big[10^{24}TrigReduce\big[meandPd\Omega2\big[\theta,\varphi,3000,310^{10}\big]\big]\big]}$

$In[23]:=$ $\mathbf{ParametricPlot3D[\{f[\theta,\varphi]\ Sin[\theta]\ Cos[\varphi],f[\theta,\varphi]\ Sin[\theta]\ Sin[\varphi],}$
$\qquad\qquad \mathbf{f[\theta,\varphi]\ Cos[\theta]\},\{\theta,0,\pi\},\{\varphi,0,2\,\pi\}]}$

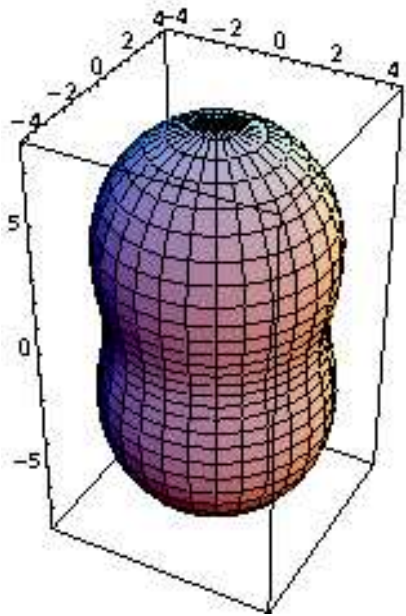

This looks exactly equal to the plot of only the dipole–contribution.

It should also be mentioned, that the $\varphi$–dependent terms originate from the electric quadrupole–momentum (and its mixed terms).

If one just looks at a rotating quadrupole (of the same value as the corresponding contribution of the Möbius strip used above), the radiation characteristics would be but with a scaling factor of $10^{-12}$ relatively to the graphics above.

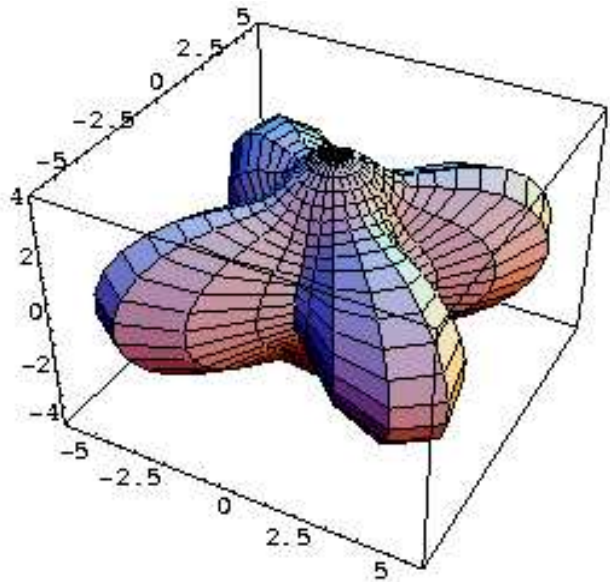

## 5.2.2 Rotation of a given Area

For the calculation of the non–conductive Möbius strip with constant area–charge (see 3.2) one again simplifies equations 5.10 to 5.12 for the special case of rotation, using the same ideas as in the previous section.

If one point on the rotation axis is $A$, one introduces a coordinate system with origin in this point. This simply results in a shift in the parametrisation vector, which is in this case taken as

`In[1]:= A = { 0,0,0 } .`

To make the rotation–axis the $z$–axis the resulting vector(s) have to be transformed with the same matrix as in 5.2.1, in this case of course again the unit matrix:

*In[2]:=* **ParamVector =**
           **B[0,0].**
             **(-A + {Cos[φ] + t Cos[φ/2] Cos[φ], Sin[φ] + t Cos[φ/2] Sin[φ],**
                **t Sin[φ/2]})**

*Out[2]=* $\left\{ \text{Cos}[\varphi] + \text{t Cos}\left[\frac{\varphi}{2}\right] \text{Cos}[\varphi], \text{Sin}[\varphi] + \text{t Cos}\left[\frac{\varphi}{2}\right] \text{Sin}[\varphi], \text{t Sin}\left[\frac{\varphi}{2}\right] \right\}$ .

The total charge is taken to be 1 which results in the (constant) area–charge

*In[3]:=* $\sigma = \dfrac{1}{6.35327}$ .

For a numerical integration of $p_0$ (the argument for pulling the rotation matrix out of the integral goes like in 5.2.1) its three components have to be calculated separately.

*In[4]:=* **pZeroX =**

         **NIntegrate$\left[\dfrac{\sigma}{2} \sqrt{4 + 3\,t^2 + 8\,t\,\text{Cos}\left[\frac{\varphi}{2}\right] + 2\,t^2\,\text{Cos}[\varphi]}\right.$**

            **$\left(\text{Cos}[\varphi] + t\,\text{Cos}\left[\frac{\varphi}{2}\right]\,\text{Cos}[\varphi]\right), \{\varphi, 0, 2\pi\}, \{t, -0.5, 0.5\}\Big]$**
*Out[4]=* 0.020587

Doing the corresponding calculation for the *y*–component there arise some failure notices:

*In[5]:=* **pZeroY =**

         **NIntegrate$\left[\dfrac{\sigma}{2} \sqrt{4 + 3\,t^2 + 8\,t\,\text{Cos}\left[\frac{\varphi}{2}\right] + 2\,t^2\,\text{Cos}[\varphi]}\right.$**

            **$\left(\text{Sin}[\varphi] + t\,\text{Cos}\left[\frac{\varphi}{2}\right]\,\text{Sin}[\varphi]\right), \{\varphi, 0, 2\pi\}, \{t, -0.5, 0.5\}\Big]$**

NIntegrate :: ncvb : NIntegrate failed to converge to prescribed accuracy after
    1 recursive bisections in *t* near {φ, t} = {φ, t}.

NIntegrate :: tmap :
   NIntegrate is unable to achieve the tolerances specified by the
        PrecisionGoal and AccuracyGoal options because the working precision is
        insufficient. Try increasing the setting of the WorkingPrecision option.

NIntegrate :: ploss :
   Numerical integration stopping due to loss of precision. Achieved
        neither the requested PrecisionGoal nor AccuracyGoal;
     suspect one of the following : highly oscillatory integrand or the true value of the
            integral is 0. If your integrand is oscillatory try using the option Method->
        Oscillatory in NIntegrate.
*Out[5]=* $3.00457 \times 10^{-15}$ .

For symmetry reasons it's clear that the result has to be zero, as well as the *z*–component (where the same messages emerge). The resulting time–dependent dipole–momentum is with

*In[6]:=* **pZero = {pZeroX, 0, 0}**

*In[7]:=* **p = Bz[ω t].pZero**
*Out[7]=* {0.020587 Cos[t ω], 0.020587 Sin[t ω], 0} .

The dipole radiation power is from that

*In[8]:=* **meandPdΩl[θ, φ, ω, c]**

$$Out[8]= \frac{0.0000168634\, \omega^4 + 0.0000168634\, \omega^4\, Cos[\theta]^2}{c^3} \, ,$$

the characteristic plot looks as follows which shows that it is about 10 times higher radiative than the conducting Möbius strip but of equal shape.

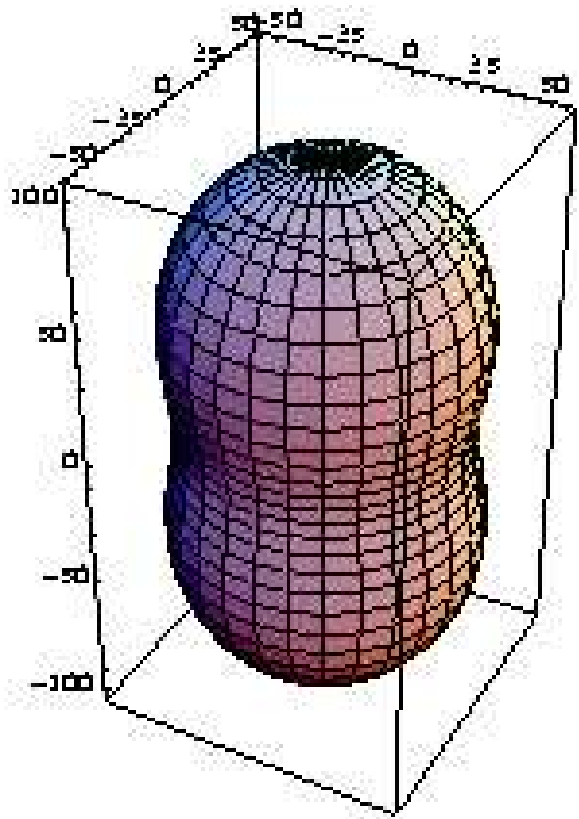

It is clear that the corrections of the higher order terms (magnetic dipole, electric quadrupole) are negligible, for completeness they are also calculated:

$In[9]:=$ **mPrimeX = NIntegrate$\left[\text{rz rx}\dfrac{\sigma}{2}\sqrt{4 + 3\,t^2 + 8\,t\,\text{Cos}\left[\frac{\varphi}{2}\right] + 2\,t^2\,\text{Cos}[\varphi]}\right.$,**

**$\{\varphi, 0, 2\,\pi\}, \{t, -0.5, 0.5\}\big]$**

NIntegrate :: ncvb :  NIntegrate failed to converge to prescribed accuracy after
      6 recursive bisections in $\varphi$ near $\{\varphi, t\} = \{\varphi, t\}$.

NIntegrate :: tmap :
   NIntegrate is unable to achieve the tolerances specified by the
        PrecisionGoal and AccuracyGoal options because the working precision is
        insufficient.  Try increasing the setting of the WorkingPrecision option.

NIntegrate :: ploss :
   Numerical integration stopping due to loss of precision. Achieved
        neither the requested PrecisionGoal nor AccuracyGoal;
      suspect one of the following :  highly oscillatory integrand or the true value of the
              integral is 0. If your integrand is oscillatory try using the option Method- >
        Oscillatory in NIntegrate.

$Out[9]=$ $-1.5992 \times 10^{-17}$

$In[10]:=$ **mPrimeY = NIntegrate$\left[\text{rz ry}\dfrac{\sigma}{2}\sqrt{4 + 3\,t^2 + 8\,t\,\text{Cos}\left[\frac{\varphi}{2}\right] + 2\,t^2\,\text{Cos}[\varphi]}\right.$,**

**$\{\varphi, 0, 2\,\pi\}, \{t, -0.5, 0.5\}\big]$**

$Out[10]=$ 0.041217

$In[11]:=$ **mPrimeZ = NIntegrate$\left[\text{rz rz}\dfrac{\sigma}{2}\sqrt{4 + 3\,t^2 + 8\,t\,\text{Cos}\left[\frac{\varphi}{2}\right] + 2\,t^2\,\text{Cos}[\varphi]}\right.$,**

**$\{\varphi, 0, 2\,\pi\}, \{t, -0.5, 0.5\}\big]$**

$Out[11]=$ 0.0420065

$In[12]:=$ **mPrime = {0, mPrimeY, mPrimeZ}**

$Out[12]=$ {0, 0.041217, 0.0420065}

$In[13]:=$ **QzeroXX =**

   **3 NIntegrate$\left[\text{rx rx}\dfrac{\sigma}{2}\sqrt{4 + 3\,t^2 + 8\,t\,\text{Cos}\left[\frac{\varphi}{2}\right] + 2\,t^2\,\text{Cos}[\varphi]}\right.$,**

   **$\{\varphi, 0, 2\,\pi\}, \{t, -0.5, 0.5\}\big]$**

$Out[13]=$ 1.68418

*In[14]:=* **QzeroXY =**

$$3 \, \mathbf{NIntegrate}\left[\mathbf{rx}\,\mathbf{ry}\,\frac{\sigma}{2}\,\sqrt{4 + 3\,t^2 + 8\,t\,\mathbf{Cos}\left[\frac{\varphi}{2}\right] + 2\,t^2\,\mathbf{Cos}[\varphi]}\,,\right.$$

$$\left.\{\varphi, 0, 2\,\pi\}, \{t, -0.5, 0.5\}\right]$$

NIntegrate :: ncvb : NIntegrate failed to converge to prescribed accuracy after
6 recursive bisections in $\varphi$ near $\{\varphi, t\} = \{\varphi, t\}$.

NIntegrate :: tmap :
  NIntegrate is unable to achieve the tolerances specified by the
      PrecisionGoal and AccuracyGoal options because the working precision is
      insufficient. Try increasing the setting of the WorkingPrecision option.

NIntegrate :: ploss :
  Numerical integration stopping due to loss of precision. Achieved
      neither the requested PrecisionGoal nor AccuracyGoal;
    suspect one of the following : highly oscillatory integrand or the true value of the
          integral is 0. If your integrand is oscillatory try using the option Method->
      Oscillatory in NIntegrate.

*Out[14]=* $4.14319 \times 10^{-16}$

*In[15]:=* **QzeroXZ =**

$$3 \, \mathbf{NIntegrate}\left[\mathbf{rx}\,\mathbf{rz}\,\frac{\sigma}{2}\,\sqrt{4 + 3\,t^2 + 8\,t\,\mathbf{Cos}\left[\frac{\varphi}{2}\right] + 2\,t^2\,\mathbf{Cos}[\varphi]}\,,\right.$$

$$\left.\{\varphi, 0, 2\,\pi\}, \{t, -0.5, 0.5\}\right]$$

NIntegrate :: ncvb : NIntegrate failed to converge to prescribed accuracy after
6 recursive bisections in $\varphi$ near $\{\varphi, t\} = \{\varphi, t\}$.

NIntegrate :: tmap :
  NIntegrate is unable to achieve the tolerances specified by the
      PrecisionGoal and AccuracyGoal options because the working precision is
      insufficient. Try increasing the setting of the WorkingPrecision option.

NIntegrate :: ploss :
  Numerical integration stopping due to loss of precision. Achieved
      neither the requested PrecisionGoal nor AccuracyGoal;
    suspect one of the following : highly oscillatory integrand or the true value of the
          integral is 0. If your integrand is oscillatory try using the option Method->
      Oscillatory in NIntegrate.

*Out[15]=* $-4.79759 \times 10^{-17}$

*In[16]:=* **QzeroYY =**

$$3\,\text{NIntegrate}\Big[\text{ry}\,\text{ry}\,\frac{\sigma}{2}\,\sqrt{4 + 3\,t^2 + 8\,t\,\text{Cos}\Big[\frac{\varphi}{2}\Big] + 2\,t^2\,\text{Cos}[\varphi]}\,,$$

$$\{\varphi, 0, 2\,\pi\}, \{t, -0.5, 0.5\}\Big]$$

*Out[16]=* 1.68418

*In[17]:=* **QzeroYZ =**

$$3\,\text{NIntegrate}\Big[\text{ry}\,\text{rz}\,\frac{\sigma}{2}\,\sqrt{4 + 3\,t^2 + 8\,t\,\text{Cos}\Big[\frac{\varphi}{2}\Big] + 2\,t^2\,\text{Cos}[\varphi]}\,,$$

$$\{\varphi, 0, 2\,\pi\}, \{t, -0.5, 0.5\}\Big]$$

*Out[17]=* 0.123651

*In[18]:=* **QzeroZZ =**

$$3\,\text{NIntegrate}\Big[\text{rz}\,\text{rz}\,\frac{\sigma}{2}\,\sqrt{4 + 3\,t^2 + 8\,t\,\text{Cos}\Big[\frac{\varphi}{2}\Big] + 2\,t^2\,\text{Cos}[\varphi]}\,,$$

$$\{\varphi, 0, 2\,\pi\}, \{t, -0.5, 0.5\}\Big]$$

*Out[18]=* 0.126019

*In[19]:=* **Qzero = {{QzeroXX, 0, 0}, {0, QzeroYY, QzeroYZ}, {0, QzeroYZ, QzeroZZ}};**
**MatrixForm[Qzero]**

*Out[19]=* $\begin{pmatrix} 1.68418 & 0 & 0 \\ 0 & 1.68418 & 0.123651 \\ 0 & 0.123651 & 0.126019 \end{pmatrix}$

*In[20]:=* **qDoubleDot =**
**Bdotdot.pZero-**
$$\frac{\omega}{2\,c}\text{Cross[Bdotdot.mPrime, {Sin[}\theta\text{] Cos[}\varphi\text{], Sin[}\theta\text{] Sin[}\varphi\text{], Cos[}\theta\text{]}}]+$$
$$\frac{1}{6\,c}\text{{Sin[}\theta\text{] Cos[}\varphi\text{], Sin[}\theta\text{] Sin[}\varphi\text{], Cos[}\theta\text{]}}.$$
$$\text{(Bdotdot.Qzero.Bz[-}\omega\text{ t] + 3 Bdotdot.Qzero.T[Bdot]+}$$
$$\text{3 Bdot.Qzero.T[Bdotdot] + Bz[}\omega\text{ t].Qzero.T[Bdotdot])}$$

*In[21]:=* **meandPd$\Omega$[$\theta$, $\varphi$, $\omega$, c]**

*Out[21]=* $\dfrac{0.0000168634\,\omega^4 + 0.0000168634\,\omega^4\,\text{Cos}[\theta]^2}{c^3}+$

$\dfrac{-0.0000675242\,\omega^5\,\text{Cos}[\theta] - 0.0000675242\,\omega^5\,\text{Cos}[\theta]^3}{c^4}+$

$\dfrac{1}{c^5}\,(-8.44935\times10^{-6}\,\omega^6 + 0.0000844935\,\omega^6\,\text{Cos}[\theta]^2+$

$1.08377\times10^{-14}\,\omega^6\,\text{Cos}[\varphi]^2 - 2.16754\times10^{-14}\,\omega^6\,\text{Cos}[\theta]^2\,\text{Cos}[\varphi]^2+$

$2.16754\times10^{-14}\,\omega^6\,\text{Cos}[\theta]^2\,\text{Cos}[\varphi]^4 + \omega^6\,\text{Cos}[\theta]^4$

$(0.0000591455 + 1.08377\times10^{-14}\,\text{Cos}[\varphi]^2 - 1.08377\times10^{-14}\,\text{Cos}[\varphi]^4)+$

$8.44935\times10^{-6}\,\omega^6\,\text{Sin}[\theta]^4 + 1.54824\times10^{-15}\,\omega^6\,\text{Cos}[\varphi]^2\,\text{Sin}[\theta]^4+$

$\omega^6\,\text{Cos}[\varphi]^4\,(-1.08377\times10^{-14} - 1.54824\times10^{-15}\,\text{Sin}[\theta]^4))$

The shape of the total radiation power remains unchanged, but the quadrupole–part has a quite interesting form:

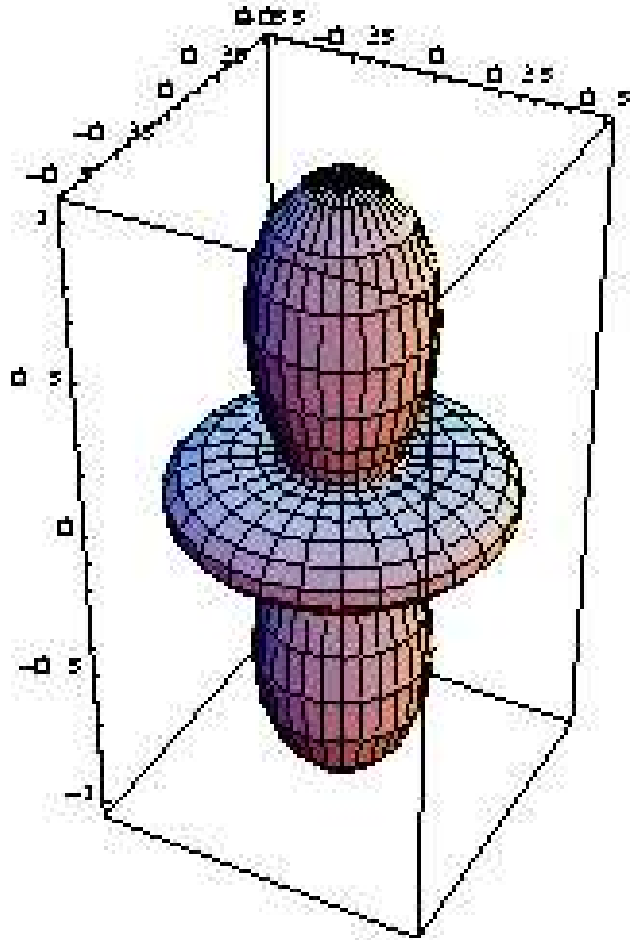

It is obvious that the rotation symmetry results from the block–diagonal form of the $Q$–matrix.

# Chapter 6

# Conclusion

It is out of question that the Möbius strip is a quite interesting two–dimensional structure which was also worthwile to consider in an electro–dynamic discussion.

The idea of sending a current around the Möbius strip to drive a dipole (electric or magnetic) in the middle of the strip was not discussed, because it is clear that this would not result in effects not realisable with an ordinary wire. Instead it was described how the Möbius strip behaves in electrostatics, conducting or non–conducting, and electrodynamics, taking a closer look at rotating charged structures.

As was seen in this work, the Möbius strip behaved quite interesting but not unexpected. Especially for rotation there cannot be any peak in the radiation scheme, because one can always do the calculation of the leading term which will then dominate and therefore determine the radiation. This can vary in magnitude but not in shape.

The benefit of this diploma work are the two programmed *Mathematica*–packages, used and described in chapters three and four. The printout of these packages in notebook–form is given in the appendix, together with the necessary C–code for the needed functions. One can also find a more detailed description of how to combine *Mathematica* and C in appendix A.

These packages can not only be used for gaining more insight in basic lectures about electrostatics. They are also applicable for visualising potentials other than the Coulomb one.

# Appendix A

# Programming *Mathematica* Functions in C

This appendix will describe the basic commands necessary to combine *Mathematica*– and C–programs. A more detailed explanation can be found in the *Mathematica* book, section 2.13.3 and references, available in [4].

The C–file itself has not to be modified very much. For a basic function there are only two things:

- The library `mathlink.h` has to be included

- The definition– and return–lines of the main function have to be

  ```
  int main(int argc, char *argv[]) {
      return MLMain(argc, argv);    } .
  ```

In the C–file set up in this way there can be programmed functions in any way one can imagine — and program in C. It can also take nearly any arguments from within *Mathematica*, which will be described immediately.

To enable *Mathematica* to use the programmed C–function a so called *MathLink* template (`.tm`–file) has to be written. A simple one looks as follows:

```
:Begin:
:Function:       c_function
:Pattern:        mma_function[x_Integer, y_Integer]
:Arguments:      {x, y,x*y}
:ArgumentTypes:  {Integer, Integer, Integer}
```

```
:ReturnType:    Integer
:End:
```

It has the obligatorial `:Begin:` and `:End:` clauses and the lines

- `:Function:`        name of the C–function

- `:Pattern:`         name of the *Mathematica*–function and its arguments

- `:Arguments:`       actual arguments given to the C–function, contained in curly brackets

- `:ArgumentTypes:`   the variable types of the C–arguments, again in brackets

- `:ReturnType:`      return type of the C–function

One can also add an arbitrary number of `:Evaluate:   <mma-code>` lines that are evaluated *once* when the function is installed (see below).

The simplest argument types are

| Argument type | C–code in function definition |
|---|---|
| Integer | int |
| Real | double |
| String | char * |
| Symbol | char* |
| IntegerList | int *, long |
| RealList | double *, long |
| Manual | void |

Mark that for lists not only a pointer of appropriate type has to be declared but also a `long` variable to pass the length of the list.

Manual can be specified if one either cannot say which return–type the function will give or if the returned arguments are not of simple type (Integer, Real; *not* IntegerList, RealList). The C–function then has to contain one of the commands

```
MLPutDouble, MLPutFunction, MLPutIntegerList, MLPutRealArray,
MLPutSymbol, ...
```

or similar commands, all of the form `MLPut<something>` (see [4]), which constitute the return value.

If the C–file and *MathLink* template are finished the *Mathematica* compiler command can be given:

```
mcc -o <installFile> <MLtemplatefile.tm> <c_code_file.c>  .
```

The programmed C–function can now be used from a *Mathematica* notebook by entering

*In[1]:=* **link = Install[" < installFile > "]**

and calling the function specified in the `.tm`–line `:Pattern:`.

The packages `AreaPotential.m` and `AreaConductor.m` use one more special feature of *Mathematica*, the command `Splice`.

If one wants to give the C–function not only values (single or in lists, arrays) but also arbitrary functions this cannot be done by simple function–arguments. In the case of the "Area–packages" this is the case for the (completely arbitrary) parametrisation of the area and the potential function.

The way to solve this problem is, to program only a template C–file, leaving out the explicit definitions of the functions mentioned above. Instead, one has to specify a "placeholder" to let mathematica fill in the missing parts. It goes like this:

Assume one has a *Mathematica*–function which one wants to use in a C–program,

*In[2]:=* $f[x\_] := \frac{x}{1 + x^2}$ .

Then one programs a template C–file `file.template` which includes the function

```
double f(double a) {
   return <*f[a]*> ;
} .
```

Now one runs the `Splice`–command

*In[3]:=* **Splice["file.template", "file.c", FormatType → CForm]** .

This command searches the file `file.template` for clauses delimited by `<*` and `*>`, executes them as if they were common *Mathematica* expressions and creates the file `file.c` with the clauses replaced by the result of the corresponding evaluation.

If this file also includes an external function one can now use

*In[4]:=* **Run["mcc - o installFile file.tm file.c"]**

(`file.tm` already has to exist) and

*In[5]:=* **Install["installFile"]**

to make the external function available, which now depends (in general) on the previously defined *Mathematica*–function f.

This principle is also used in the packages listed below.

# Appendix B

# *Mathematica* Packages

## B.1    AreaPotential.m

*In[1]:=* **BeginPackage["`AreaPotential`"]**

## Usage

*In[2]:=* **AreaPotential :: usage =**
          **"contains : \n SetArea \n SetDistribution \n SetPotentialLaw**
              **\n GridValues\n Potential\n ViewPotential\n AnimatePotential"**

*In[3]:=* **SetArea :: usage =**
          **"SetArea[ParamVector, aRange, bRange, accuracy]\n Sets new**
              **Area through parametrisation\n\nParamVector is a List of**
              **three Elements that have to depend on two parameters**
              **which have to be the symbols ′a′ and ′b′\n\nxRange is a**
              **List of one Symbol (′a′ or ′b′) and two Real values (lower**
              **and upper parameter - bound)\n\naccuarcy specifys the density**
              **of the calculation - grid on the area, its default - value is 30"**

*In[4]:=* **SetDistribution :: usage =**
          **"SetDistribution[chargeFunction]\n Sets the charge distribution**
              **function on the surface\n\nchargeFunction is an expression**
              **which may only consist of a and b"**

*In[5]:=* `SetPotentialLaw :: usage =`
        `"SetPotentialLaw[PotLaw]\n Changes the potential law of ´`
            `point charges´, default is the Coulomb - law\n\nPotLaw is`
            `an expression that must only contain the symbols ´x´, ´y´`
            `and ´z´ (the cartesian coordinates, respectively)"`

*In[6]:=* `GridValues :: usage =`
        `"GridValues[xRange, yRange, zRange, accuracy, maxPotential]\n`
            `Returns a grid of potential - values\n\n_Range is a list`
            `of two values, lower and upper bound to specify the ashlar`
            `of interest\n\naccuracy specifies the density of the grid`
            `\n\nWith no arguments specified GridValues returns the`
            `grid of the last calculation"`

*In[7]:=* `Potential :: usage =`
        `"Potential[{x, y, z}]\n Returns the electrostatic potential`
            `at the point (x, y, z) based on the previous grid - calculation"`

*In[8]:=* `ViewPotential :: usage =`
        `"ViewPotential[planeNumber, opt___] OR\nViewPotential[`
            `planeNumber, Grid, opt___]\n Returns a graphics object`
            `showing the surface cut by the plane of interest and the`
            `potential on this plane.\n\nIf ´Grid´ is left out the grid`
            `of the last calculation is taken; \n´opt___´ stands for an`
            `optional argument describing to which axes the plane of`
            `interest should be parallel, e.g. ´plane → xy´ (which is default)"`

*In[9]:=* `AnimatePotential :: usage =`
        `"PotentialAnimation[opt___]\n Returns a bunch of graphics`
            `which are animated when double - clicked on the first one."`

Debugging

*In[10]:=* `Parameters :: usage = ""`

*In[11]:=* `Begin["`Private`"]`

*In[12]:=* `Off[General :: "spell1"]`

# Setting default values for variables needed in computation (all unknown to user)

Charge distribution on the surface

*In[13]:=* `chargeDist = 1.`

Parametrisation-vector of the area, parameters are (always) a and b, set to Möbius strip:

```
In[14]:= ParamVector = {Cos[b] + a Cos[b/2] Cos[b], Sin[b] + a Cos[b/2] Sin[b],
            a Sin[b/2]}
```

Tangent-vector in the direction of the a-parametrisation-curve:

```
In[15]:= av = Simplify[D[ParamVector, a]]
```

Tangent-vector in the direction of the b-parametrisation-curve:

```
In[16]:= bv = D[ParamVector, b]
```

Normal-vector to the area:

```
In[17]:= nv = Simplify[Cross[av, bv]]
```

Infinitesimal area-element:

```
In[18]:= dO = Simplify[chargeDist Sqrt[nv.nv]]
```

Split parametrisation-vector into 3 functions

```
In[19]:= xParam[a_, b_] := Evaluate[ParamVector[[1]]]
```

```
In[20]:= yParam[a_, b_] := Evaluate[ParamVector[[2]]]
```

```
In[21]:= zParam[a_, b_] := Evaluate[ParamVector[[3]]]
```

Parameter for the accuracy of the integral-calculation:

```
In[22]:= acc = 30
```

1/distance is default for the (electrostatic) potential law:

```
In[23]:= PotentialLaw = 1./Sqrt[x * x + y * y + z * z]
```

Boundaries for the parameters a/b and the ashlar of interest:

```
In[24]:= aLower = -0.5; aUpper = 0.5;

        bLower = 0.; bUpper = 2 N[π];

        xLower = -3.; xUpper = 3.;

        yLower = -3.; yUpper = 3.;

        zLower = -3.;
        zUpper = 3.;
```

The three lengths of the ashlar are divided into 'accuracy' parts to constitute the grid:

```
In[25]:= accuracy = 25
```

The upper boundary for the potential, necessary because of the incontinuous calculation, is

```
In[26]:= maxPotential = 8.
```

Variable to store the outcome of the potential calculation (especially if user doesn't store it for himself):

```
In[27]:= InternalGridDefault = "Not yet calculated, use 'GridValues'."
```

## Internal Functions

### The function, that writes the parameters into the C-file:

```
In[28]:= mySettings[chDist_, ParamVec_, al_, au_, bl_, bu_, accuracy_,
            PotLaw_] :=
         (ParamVector = ParamVec; chargeDist = chDist;
          xParam[x_, y_] := Evaluate[ParamVec[[1]] /. a → x /. b → y];
          yParam[x_, y_] := Evaluate[ParamVec[[2]] /. a → x /. b → y];
          zParam[x_, y_] := Evaluate[ParamVec[[3]] /. a → x /. b → y];
          av = Simplify[D[ParamVector, a]]; bv = D[ParamVector, b];
          nv = Simplify[Cross[av, bv]]; dO = Simplify[chargeDist Sqrt[nv.nv]];
          aLower = al; aUpper = au; bLower = bl; bUpper = bu;
          acc = accuracy; PotentialLaw = PotLaw;
          InternalGrid = InternalGridDefault; )
```

### Find out, whether a vector is a valid parametrisation-vector (3-dim., only depends on a and b):

```
In[29]:= ParamQ[{xv_, yv_, zv_}] :=
            Module[{var}, var = xv + yv + zv /. Global`a → 1. /. Global`b → 1.;
             MachineNumberQ[var]]
```

### Find out, whether an expression is a valid charge distribution function (only depends on a and b):

```
In[30]:= distQ[expression_] :=
            Module[{var}, var = expression /. Global`a → 1. /. Global`b → 1.;
             MachineNumberQ[var]]
```

**Find out, whether an expression is a valid potential law (depends only on x, y and z):**

```
In[31]:= LawQ[arg_] :=
            Module[{var}, var = arg/. Global`x → 1. /. Global`y → 1. /. Global`z → 1.;
              MachineNumberQ[var]]
```

**Find out, whether an expression is a valid range-list for the arguments of GridValues:**

```
In[32]:= RangeQ[arg_] := Module[{argument}, argument = Map[N, arg];
            If[Length[argument] == 2, VectorQ[argument, NumberQ], False]]
```

**Messages:**

```
In[33]:= GridValues :: "CProgram" =
            "Compilation and installation successful, starting calculation."
```

# User-Functions

**User-function for setting (only) the area of the charged thing:**

```
In[34]:= SetArea[ParamVec_?ParamQ, {Global`a, al_Real, au_Real},
            {Global`b, bl_Real, bu_Real}, accuracy_Integer :30] :=
            mySettings[chargeDist, ParamVec/. Global`a → a/.Global`b → b,
             al, au, bl, bu, accuracy, PotentialLaw]
```

**User-function for setting (only) the charge distribution function:**

```
In[35]:= SetDistribution[expression_] :=
            mySettings[expression/. Global`a → a/.Global`b → b, ParamVector,
             aLower, aUpper, bLower, bUpper, acc, PotentialLaw]
```

**User-function for setting (only) the potential law:**

```
In[36]:= SetPotentialLaw[PotLaw_?LawQ] :=
            mySettings[chargeDist, ParamVector, aLower, aUpper, bLower,
             bUpper, acc, PotLaw]
```

**User-function to get the potential-values on the specified grid:**

```
In[37]:= GridValues[xRange_?RangeQ, yRange_?RangeQ, zRange_?RangeQ,
            gotAccuracy_Integer :accuracy, gotMaxPotential_Real :maxPotential] :=
         (If[xRange[[1]] < xRange[[2]],
             xLower = xRange[[1]]; xUpper = xRange[[2]],
             xLower = xRange[[2]]; xUpper = xRange[[1]]];

          If[yRange[[1]] < yRange[[2]],
             yLower = yRange[[1]]; yUpper = yRange[[2]],
             yLower = yRange[[2]]; yUpper = yRange[[1]]];

          If[zRange[[1]] < zRange[[2]],
             zLower = zRange[[1]]; zUpper = zRange[[2]],
             zLower = zRange[[2]]; zUpper = zRange[[1]]];

          accuracy = gotAccuracy; maxPotential = gotMaxPotential;

          Uninstall[link];
          Splice["potential.template", "potential.c", FormatType → CForm];
          Run["mcc - o potential potential.tm potential.c"];
          link = Install["potential"];
          Message[GridValues :: "CProgram"];

          InternalGrid = Global`GridVals[xRange, yRange, zRange, accuracy,
             maxPotential])

In[38]:= GridValues[] := InternalGrid
```

**User-function to get the potential at an arbitrary position based on previous GridValues**

```
In[39]:= Potential :: "position" = "`1` = `2` out of range (`3`, `4`)."

In[40]:= ValidX[x_] := If[x < xLower || x > xUpper,
            Message[Potential :: "position", "x", x, xLower, xUpper], True]

In[41]:= ValidY[y_] := If[y < yLower || y > yUpper,
            Message[Potential :: "position", "y", y, yLower, yUpper], True]

In[42]:= ValidZ[z_] := If[z < zLower || z > zUpper,
            Message[Potential :: "position", "z", z, zLower, zUpper], True]
```

```
In[43]:= Potential[{x_?ValidX, y_?ValidY, z_?ValidZ}] :=
```

$$\text{Module}\Big[\Big\{\Delta x = \frac{\text{xUpper} - \text{xLower}}{\text{accuracy}}, \Delta y = \frac{\text{yUpper} - \text{yLower}}{\text{accuracy}}, \Delta z = \frac{\text{zUpper} - \text{zLower}}{\text{accuracy}},$$

$$\xi, yi, zi, dx, dy, dz\Big\}, \xi = \text{IntegerPart}\Big[\frac{x - \text{xLower}}{\Delta x}\Big] + 1;$$

$$yi = \text{IntegerPart}\Big[\frac{y - \text{yLower}}{\Delta y}\Big] + 1; zi = \text{IntegerPart}\Big[\frac{z - \text{zLower}}{\Delta z}\Big] + 1;$$

$$dx = x - \text{xLower} - (\xi - 1)\,\Delta x; dy = y - \text{yLower} - (yi - 1)\,\Delta y;$$

$$dz = z - \text{zLower} - (zi - 1)\,\Delta z;$$

$$\text{InternalGrid}[[\xi, yi, zi]]\,\frac{\Delta x - dx}{\Delta x}\frac{\Delta y - dy}{\Delta y}\frac{\Delta z - dz}{\Delta z} +$$

$$\text{InternalGrid}[[\xi + 1, yi, zi]]\,\frac{dx}{\Delta x}\frac{\Delta y - dy}{\Delta y}\frac{\Delta z - dz}{\Delta z} +$$

$$\text{InternalGrid}[[\xi, yi + 1, zi]]\,\frac{\Delta x - dx}{\Delta x}\frac{dy}{\Delta y}\frac{\Delta z - dz}{\Delta z} +$$

$$\text{InternalGrid}[[\xi + 1, yi + 1, zi]]\,\frac{dx}{\Delta x}\frac{dy}{\Delta y}\frac{\Delta z - dz}{\Delta z} +$$

$$\text{InternalGrid}[[\xi, yi, zi + 1]]\,\frac{\Delta x - dx}{\Delta x}\frac{\Delta y - dy}{\Delta y}\frac{dz}{\Delta z} +$$

$$\text{InternalGrid}[[\xi + 1, yi, zi + 1]]\,\frac{dx}{\Delta x}\frac{\Delta y - dy}{\Delta y}\frac{dz}{\Delta z} +$$

$$\text{InternalGrid}[[\xi, yi + 1, zi + 1]]\,\frac{\Delta x - dx}{\Delta x}\frac{dy}{\Delta y}\frac{dz}{\Delta z} +$$

$$\text{InternalGrid}[[\xi + 1, yi + 1, zi + 1]]\,\frac{dx}{\Delta x}\frac{dy}{\Delta y}\frac{dz}{\Delta z}\Big]$$

## Initialisation

**Write the default-values to C-file, compile it and link the executable into Mathematica:**

```
In[44]:= mySettings[chargeDist, ParamVector, aLower, aUpper, bLower,
            bUpper, acc, PotentialLaw];

        Splice["potential.template", "potential.c", FormatType -> CForm];

        Run["mcc - o potential potential.tm potential.c"];

        link = Install["potential"]
```

## Graphical Functions

### Extraction of desired elements of a grid

*In[45]:=* `XYPlane[g_, zi_] := Map[Map[Part[#, zi] &, #] &, g]`

*In[46]:=* `XZPlane[g_, yi_] := Map[Part[#, yi] &, g]`

*In[47]:=* `YZPlane[g_, xi_] := Part[g, ξ]`

### Plotting of the slicing plane

*In[48]:=* `PlaneXY[zi_] :=`
            `ParametricPlot3D[{x, y, zLower + (zi - 1) (zUpper - zLower)/accuracy},`
               `{x, xLower, xUpper}, {y, yLower, yUpper}, DisplayFunction → Identity];`

*In[49]:=* `PlaneXZ[yi_] :=`
            `ParametricPlot3D[{x, yLower + (yi - 1) (yUpper - yLower)/accuracy, z},`
               `{x, xLower, xUpper}, {z, zLower, zUpper}, DisplayFunction → Identity];`

*In[50]:=* `PlaneYZ[xi_] :=`
            `ParametricPlot3D[{xLower + (ξ - 1) (xUpper - xLower)/accuracy, y, z},`
               `{y, yLower, yUpper}, {z, zLower, zUpper}, DisplayFunction → Identity];`

### Viewing slicing plane and parametrized area from specified viewing point

pool for viewing-point-options

*In[51]:=* `views = {{{-2.4, 1.3, -2}, {0, 0, -1}}, {{-2.4, 2, 1.3}, {0, 1, 0}},`
            `{{-2, -2.4, 1.3}, {-1, 0, 0}}}`

*In[52]:=* `viewPlane[i_, plane_] :=`
            `Show[ParametricPlot3D[{xParam[t, φ], yParam[t, φ], zParam[t, φ]},`
               `{t, aLower, aUpper}, {φ, bLower, bUpper},`
               `AxesLabel → {Global`x, Global`y, Global`z}, Boxed → False,`
               `SphericalRegion → True,`
               `PlotRange → {zLower + zLower/100, zUpper + zUpper/100},`
               `DisplayFunction → Identity,`
               `ViewPoint →`
                 `views[[First[First[Position[PlaneXY PlaneXZ PlaneYZ, plane]]]]][[`
                   `1]],`
               `ViewVertical →`
                 `views[[First[First[Position[PlaneXY PlaneXZ PlaneYZ, plane]]]]][[`
                   `2]], plane[i]];`

**Main user-function for viewing the Potential**

```
In[53]:= PlaneRangeQ[number_] := number > 0 && number < accuracy + 2

In[54]:= optQ[opt_] := Switch[Global`plane /. {opt},
            Global`xy, True,
            Global`xz, True,
            Global`yz, True,
            _, False]

In[55]:= Options[ViewPotential] = {Global`plane → Global`xy}

In[56]:= ViewPotential[planeNumber_Integer?PlaneRangeQ, myGrid_List,
            opt___?optQ] :=
          intermediate[Global`plane/.{opt}/.Options[ViewPotential],
            planeNumber, myGrid]

In[57]:= ViewPotential[planeNumber_Integer?PlaneRangeQ, opt___?optQ] :=
          intermediate[Global`plane/.{opt}/.Options[ViewPotential],
            planeNumber, InternalGrid]

In[58]:= intermediate[plane_, n_, myGrid_] := Switch[plane,
            Global`xy, myGraph[XYPlane, PlaneXY, n, myGrid],
            Global`xz, myGraph[XZPlane, PlaneXZ, n, myGrid],
            Global`yz, myGraph[YZPlane, PlaneYZ, n, myGrid]]

In[59]:= myGraph[gridPlane_, plane_, n_, myGrid_] :=
            Show[
              Graphics[
                {Rectangle[{0, 0}, {4, 4}, ListPlot3D[gridPlane[myGrid, n],
                    Axes → False, Boxed → False, SphericalRegion → True,
                    PlotRange → {0, maxPotential}, DisplayFunction → Identity]],
                 Rectangle[{3.8, 1.8}, {6, 4},
                    viewPlane[accuracy + 1 - Abs[n - accuracy - 1], plane]]},
              ImageSize → 700]];
```

**Animation of Potential**

```
In[60]:= AnimatePotential[opt___?optQ] := ShowAnimation[
            Table[
              ViewPotential[accuracy + 1 - Abs[n - accuracy - 1], opt],
              {n, 1, 2 accuracy - 1, 1}
            ]]
```

## Debugging

```
In[61]:= Parameters := Sequence["chargeDist = ",chargeDist,"\nParamVector = ",
            ParamVector,"\ndO = ",dO,"\nacc = ",acc,"\nPotentialLaw = ",
            PotentialLaw,"\naLower = ",aLower,", aUpper = ",aUpper,
            "\nbLower = ",bLower,", bUpper = ",bUpper,"\nxLower = ",
            xLower,", xUpper = ",xUpper,"\nyLower = ",yLower,", yUpper = ",
            yUpper,"\nzLower = ",zLower,", zUpper = ",zUpper,"\naccuracy = ",
            accuracy,", maxPotential = ",maxPotential,"\nInternalGrid = ",
            InternalGrid]
```

## End of Package

```
In[62]:= End[]
```

```
In[63]:= EndPackage[]
```

# B.2    AreaConductor.m

```
In[1]:= BeginPackage["`AreaConductor`"]
```

## Usage

```
In[2]:= AreaConductor :: usage =
            "contains : \n SetArea \n SetPotentialLaw\n ChargePositions
            \n Potential\n SetCharges"
```

```
In[3]:= SetArea :: usage =
            "SetArea[ParamVector,aRange,bRange]\n Sets new Area through
            parametrisation\n\nParamVector is a List of three Elements
            that have to depend on two parameters which have to be
            the symbols 'a' and 'b'\n\nxRange is a List of one Symbol ('
            a' or 'b') and two Real values (lower and upper parameter - bound)"
```

```
In[4]:= SetPotentialLaw :: usage =
            "SetPotentialLaw[PotLaw]\n Changes the potential law of '
            point charges', default is the Coulomb - law\n\nPotLaw is
            an expression that must only contain the symbol 'r' (the
            distance from a point charge, respectively)"
```

```
In[5]:= ChargePositions :: usage =
          "ChargePositions[number, steps, [speed]]\nReturns the positions
            of 'number' point - charges after 'steps' movements towards
            equilibrium position; 'speed' is an optional Real number.
            \n\nOR\nChargePositions[pos2D, steps, [speed]]\nsimilar
            to above, but the first argument is the list of positions
            of point - charges on the surface"
```

```
In[6]:= Potential :: usage =
          "Potential[{x, y, z}]\nReturns the Potential at position (x, y, z)"
```

```
In[7]:= SetCharges :: usage =
          "SetCharges[list]\nSets the charges on the points specified
            in 'list' for potential - calculations"
```

Debugging

```
In[8]:= Parameters :: usage = ""
```

```
In[9]:= Begin["'Private'"]
```

```
In[10]:= Off[General :: "spell1"]
```

## Setting default values for variables needed in computation (all unknown to user)

Parametrisation-vector of the area, parameters are (always) a and b, set to Möbius strip:

```
In[11]:= ParamVector = {Cos[b] + a Cos[b/2] Cos[b], Sin[b] + a Cos[b/2] Sin[b],
            a Sin[b/2]}
```

Tangent-vector in the direction of the a-parametrisation-curve:

```
In[12]:= av = Simplify[D[ParamVector, a]]
```

Tangent-vector in the direction of the b-parametrisation-curve:

```
In[13]:= bv = D[ParamVector, b]
```

Split parametrisation-vector into 3 functions

```
In[14]:= xParam[a_, b_] := Evaluate[ParamVector[[1]]]
```

```
In[15]:= yParam[a_, b_] := Evaluate[ParamVector[[2]]]
```

```
In[16]:= zParam[a_, b_] := Evaluate[ParamVector[[3]]]
```

1/distance is default for the (electrostatic) potential law:

```
In[17]:= PotentialLaw = 1./r;

         ForceLaw = D[-PotentialLaw, r];

         Law[0.] := ∞; Law[p_] := 1./p;

         cartesianLaw[{x_, y_, z_}] := Law[Sqrt[x² + y² + z²]];
```

Boundaries for the parameters a/b and the ashlar of interest:

```
In[18]:= aLower = -0.5; aUpper = 0.5;

         bLower = 0.; bUpper = 2 N[π];
```

The total charge is devided at maximum into 'MaxCharges' point charges that can move on the surface:

```
In[19]:= MaxCharges = 5000
```

Variable to store the outcome of the potential calculation (especially if user doesn't store it for himself):

```
In[20]:= InternalChargesDefault = "Notyetcalculated, use'ChargePositions'."
```

# Internal Functions

## The function, that writes the parameters into the C-file:

```
In[21]:= mySettings[ParamVec_, al_, au_, bl_, bu_, PotLaw_] :=
            (ParamVector = ParamVec;
             xParam[x_, y_] := Evaluate[ParamVec[[1]] /. a → x /. b → y];
             yParam[x_, y_] := Evaluate[ParamVec[[2]] /. a → x /. b → y];
             zParam[x_, y_] := Evaluate[ParamVec[[3]] /. a → x /. b → y];
             av = Simplify[D[ParamVector, a]]; bv = D[ParamVector, b];
             nv = Simplify[Cross[av, bv]];
             aLower = al; aUpper = au; bLower = bl; bUpper = bu;
             PotentialLaw = PotLaw; ForceLaw = D[-PotentialLaw, r];
             Law[p_] := Evaluate[PotLaw /. r → p];
             InternalCharges = InternalChargesDefault;
             Splice["conductor.template", "conductor.c", FormatType → CForm])
```

**Find out, whether a vector is a valid parametrisation-vector (3-dim., only depends on a and b):**

```
In[22]:= ParamQ[{xv_, yv_, zv_}] :=
            Module[{var}, var = xv + yv + zv /. Global`a → 1. /. Global`b → 1.;
              MachineNumberQ[var]]
```

**Find out, whether an expression is a valid potential law (depends only on x, y and z):**

```
In[23]:= LawQ[arg_] := Module[{var}, var = arg /. Global`r → 1. ;
            MachineNumberQ[var]]
```

**Partition of the charges on the parameter-area:**

```
In[24]:= partition :: "overflow" =
            "Number of charges (`1`) too high, replaced by `2`."
```

```
In[25]:= partition[zahl_Integer] :=
            Sequence @@
              Module[{z, a, b, c},
                z = If[zahl > MaxCharges, Message[partition :: "overflow", zahl,
                  MaxCharges]; MaxCharges, zahl]; a = IntegerPart[Sqrt[z]];
                c = z - a^2; If[c ≤ a, b = a, b = a + 1; c = c - a]; {a, b, c}]
```

**Messages:**

```
In[26]:= ChargePositions :: "now" =
            "positions calculated, generating potential function"
```

# User-Functions

**User-function for setting (only) the area of the charged thing:**

```
In[27]:= SetArea[ParamVec_?ParamQ, {Global`a, al_Real, au_Real},
            {Global`b, bl_Real, bu_Real}] :=
            mySettings[ParamVec /. Global`a → a /. Global`b → b, al, au, bl,
              bu, PotentialLaw]
```

**User-function for setting (only) the potential law:**

```
In[28]:= SetPotentialLaw[PotLaw_?LawQ] := (Clear[Law];
            mySettings[ParamVector, aLower, aUpper, bLower, bUpper, PotLaw])

In[29]:= SetPotentialLaw[PotLaw_?LawQ, ZeroValue_] := (Clear[Law];
            mySettings[ParamVector, aLower, aUpper, bLower, bUpper, PotLaw];
            Law[0.] := Evaluate[ZeroValue])
```

**User-function to get the positions of the point-charges:**

```
In[30]:= ChargePositions[number_Integer, steps_Integer :20, factor_Real :0.5] :=
        (Uninstall[link];

            Run["mcc - o conductor conductor.tm conductor.c"];
            link = Install["conductor"];

            temp = Global`PosFromNumber[partition[Abs[number]], Abs[steps],
                factor];
            InternalCharges = temp[[1]];

            CalculatedPotential[{x_, y_, z_}] := Evaluate[
                1/Length[InternalCharges]
                    Plus@@ ( (cartesianLaw[# - {x, y, z}]) & /@ InternalCharges)];
            temp)

In[31]:= ChargePositions[pos2D_List, steps_Integer :20, factor_Real :0.5] :=
        (Uninstall[link];

        Run["mcc - o conductor conductor.tm conductor.c"];
        link = Install["conductor"];

            temp = Global`PosFromArray[Flatten[pos2D], Abs[steps], factor];
            InternalCharges = temp[[1]];

            CalculatedPotential[{x_, y_, z_}] := Evaluate[
                1/Length[InternalCharges]
                    Plus@@ ( (cartesianLaw[# - {x, y, z}]) & /@ InternalCharges)];
            temp)

In[32]:= ChargePositions[] := InternalCharges
```

**User-function to set the 3D-positions of the point-charges:**

```
In[33]:= SetCharges[list_] :=
             (InternalCharges = list;
               CalculatedPotential[{x_, y_, z_}] :=
                 Evaluate[
                   1/Length[list] Plus@@( (cartesianLaw[# - {x, y, z}]) & /@ list) ])
```

**Potential at one point:**

```
In[34]:= Potential[point_] := CalculatedPotential[point]
```

## Initialisation

**Write the default-values to C-file, compile it and link the executable into Mathematica:**

```
In[35]:= mySettings[ParamVector, aLower, aUpper, bLower, bUpper, PotentialLaw];

         Run["mcc - o conductor conductor.tm conductor.c"];

         link = Install["conductor"]
```

## Debugging

```
In[36]:= Parameters := Sequence["ParamVector = ", ParamVector, "\nd0 = ",
             d0, "\nPotentialLaw = ", PotentialLaw, "\naLower = ", aLower,
             ", aUpper = ", aUpper, "\nbLower = ", bLower, ", bUpper = ",
             bUpper, "\nInternalCharges = ", InternalCharges]
```

## End of Package

```
In[37]:= End[]
```

```
In[38]:= EndPackage[]
```

# Appendix C

# .tm Files

For the printout of the following files the `verbatim`–environment was used (see
[3]).

## C.1    potential.tm

```
:Begin:
:Function:      GridVals
:Pattern:       GridVals[x_List, y_List, z_List, i_Integer, max_Real]
:Arguments:     {x, y, z, i, max}
:ArgumentTypes: {RealList,RealList,RealList,Integer,Real}
:ReturnType:    Manual
:End:
```

## C.2    conductor.tm

```
:Begin:
:Function: PosFromNumber
:Pattern:       PosFromNumber[numberA_Integer,numberB_Integer,
                          numberLast_Integer,steps_Integer,factor_Real]
:Arguments:     {numberA,numberB,numberLast,steps,factor}
:ArgumentTypes: {Integer,Integer,Integer,Integer,Real}
:ReturnType:    Manual
:End:
:Begin:
:Function: PosFromArray
:Pattern: PosFromArray[pos2D_List,steps_Integer,factor_Real]
:Arguments: {pos2D,steps,factor}
:ArgumentTypes: {RealList,Integer,Real}
```

```
:ReturnType: Manual
:End:
```

# Appendix D

# C Programs

## D.1  potential.template

```
#include "mathlink.h"
#include <math.h>
#include <stdlib.h>
#include <stdio.h>
#include <bits/nan.h>
#define E exp(1)
#define Power pow
#define Cos cos
#define Sin sin
#define Tan tan
#define Sqrt sqrt
#define Log log

float a0, a1, da, b0, b1, db;
float *grid;
int AreaGrid;

/**************************************
Parametrisierung der Fläche:
**************************************/

float xParam(float a,float b)
{ return <*Global`AreaPotential`Private`ParamVector[[1]]
    /. Global`AreaPotential`Private`b->b /. Global`AreaPotential`Private`a->a*>; }

float yParam(float a, float b)
{ return <*Global`AreaPotential`Private`ParamVector[[2]]
    /. Global`AreaPotential`Private`b->b /. Global`AreaPotential`Private`a->a*>; }
```

```
float zParam(float a, float b)
{ return <*Global`AreaPotential`Private`ParamVector[[3]]
    /. Global`AreaPotential`Private`b->b /. Global`AreaPotential`Private`a->a*>; }

/**************************************
Betrag des Normalvektors:
**************************************/

float dO(float a, float b)
{
  return <*Global`AreaPotential`Private`dO
    /. Global`AreaPotential`Private`b->b /. Global`AreaPotential`Private`a->a*>;
}

/**************************************
Parameterbereich festlegen:
**************************************/

void SetParams(float *a_lower, float *a_upper,
               float *b_lower, float *b_upper, int *accuracy)
{
  *a_lower=<*Global`AreaPotential`Private`aLower*>;
  *a_upper=<*Global`AreaPotential`Private`aUpper*>;
  *b_lower=<*Global`AreaPotential`Private`bLower*>;
  *b_upper=<*Global`AreaPotential`Private`bUpper*>;
  *accuracy=<*Global`AreaPotential`Private`acc*>;
}

/**************************************
Potentialgesetz:
**************************************/

float law(float x, float y, float z)
{
  return <*Global`AreaPotential`Private`PotentialLaw
    /. Global`AreaPotential`Private`x->x /. Global`AreaPotential`Private`y->y
    /. Global`AreaPotential`Private`z->z*>;
}

////////////////////////////////////////////////////////////////////////

float Integrand(float a, float b, float x, float y, float z)
{
  return dO(a,b)*law(x-xParam(a,b),y-yParam(a,b),z-zParam(a,b));
}

////////////////////////////////////////////////////////////////////////
```

```
double Potential(float x, float y, float z)
{
  float val=0.;
  int i, j;
  for(i=0;i<AreaGrid+1;i++)
  {
    for (j=0;j<AreaGrid+1;j++)
    {
      val+=Integrand(a0+(float)i*da,b0+(float)j*db,x,y,z);
    }
  }
  val=val*(a1-a0)*(b1-b0)/(float)((AreaGrid+1)*(AreaGrid+1));
  return (double)val;
}

//////////////////////////////////////////////////////////////////////

void GridVals(double *xrange, long xlen, double *yrange, long ylen,
              double *zrange, long zlen, int MainGrid, double MAX)
{
  float xmin,xmax,dx,ymin,ymax,dy,zmin,zmax,dz;
  double vals[MainGrid+1][MainGrid+1][MainGrid+1], *dummy;
  int i,j,k;
  long dims[]={MainGrid+1,MainGrid+1,MainGrid+1};

  if ((xlen != 2) || (ylen!=2) || (zlen!=2))
  {
    MLEvaluateString(stdlink,
      "Print[\"Pattern: {x_min,x_max},{y_min,y_max},{z_min,z_max},stepNumber\"]");
    return;
  }
  if((xmin=(float)xrange[0])>(xmax=(float)xrange[1]))
    {xmin=(float)xrange[1];xmax=(float)xrange[0];}
  if((ymin=(float)yrange[0])>(ymax=(float)yrange[1]))
    {ymin=(float)yrange[1];ymax=(float)yrange[0];}
  if((zmin=(float)zrange[0])>(zmax=(float)zrange[1]))
    {zmin=(float)zrange[1];zmax=(float)zrange[0];}
  dx=(xmax-xmin)/(float)MainGrid;
  dy=(ymax-ymin)/(float)MainGrid;
  dz=(zmax-zmin)/(float)MainGrid;
  for (i=0;i<=MainGrid;i++)
  {
    for (j=0;j<=MainGrid;j++)
    {
      for (k=0;k<=MainGrid;k++)
      {
        vals[i][j][k]=Potential(xmin+(float)i*dx,ymin+(float)j*dy,zmin+(float)k*dz);
if ((vals[i][j][k]>MAX) || (vals[i][j][k]==NAN)) {vals[i][j][k]=MAX;}
else if (vals[i][j][k]<-1*MAX) {vals[i][j][k]=-1*MAX;}
```

```
      }
    }
  }
  dummy=(double*)vals;
//  MLPutRealArray(stdlink, vals, dims, NULL, 3);
  MLPutRealArray(stdlink, dummy, dims, NULL, 3);
  return;
}

////////////////////////////////////////////////////////////////////////

////////////////////////////////////////////////////////////////////////

int main(int argc, char *argv[])
{
  SetParams(&a0, &a1, &b0, &b1, &AreaGrid);
  da=(a1-a0)/(float)AreaGrid; db=(b1-b0)/(float)AreaGrid;
  return MLMain(argc, argv);
}
```

## D.2   conductor.template

```
#include "mathlink.h"
#include <math.h>
#include <stdlib.h>
#include <stdio.h>
#include <bits/nan.h>
#define E exp(1)
#define Power pow
#define Cos cos
#define Sin sin
#define Tan tan
#define Sqrt sqrt
#define Log log

float a0, a1, da, b0, b1, db, multiplicator;
float pos3D[10000][3], pos2D[10000][2];
double *charges_pointer;

/**************************************
Parametrisierung der Fläche:
**************************************/

float xParam(float a,float b)
{ return <*Global`AreaConductor`Private`ParamVector[[1]]
  /. Global`AreaConductor`Private`b->b /. Global`AreaConductor`Private`a->a*>; }
```

```
float yParam(float a, float b)
{ return <*Global`AreaConductor`Private`ParamVector[[2]]
  /. Global`AreaConductor`Private`b->b /. Global`AreaConductor`Private`a->a*>; }

float zParam(float a, float b)
{ return <*Global`AreaConductor`Private`ParamVector[[3]]
  /. Global`AreaConductor`Private`b->b /. Global`AreaConductor`Private`a->a*>; }

/**************************************
Tangentialvektoren der Fläche:
**************************************/

void get_tangent_vectors(float *aVector, float *bVector, float a, float b){

    aVector[0]=<*Global`AreaConductor`Private`av[[1]]
      /. Global`AreaConductor`Private`b->b /. Global`AreaConductor`Private`a->a*>;
    aVector[1]=<*Global`AreaConductor`Private`av[[2]]
      /. Global`AreaConductor`Private`b->b /. Global`AreaConductor`Private`a->a*>;
    aVector[2]=<*Global`AreaConductor`Private`av[[3]]
      /. Global`AreaConductor`Private`b->b /. Global`AreaConductor`Private`a->a*>;
    bVector[0]=<*Global`AreaConductor`Private`bv[[1]]
      /. Global`AreaConductor`Private`b->b /. Global`AreaConductor`Private`a->a*>;
    bVector[1]=<*Global`AreaConductor`Private`bv[[2]]
      /. Global`AreaConductor`Private`b->b /. Global`AreaConductor`Private`a->a*>;
    bVector[2]=<*Global`AreaConductor`Private`bv[[3]]
      /. Global`AreaConductor`Private`b->b /. Global`AreaConductor`Private`a->a*>;
}

/**************************************
Parameterbereich festlegen:
**************************************/

void SetParams(float *a_lower, float *a_upper, float *b_lower, float *b_upper)
{
  *a_lower=<*Global`AreaConductor`Private`aLower*>;
  *a_upper=<*Global`AreaConductor`Private`aUpper*>;
  *b_lower=<*Global`AreaConductor`Private`bLower*>;
  *b_upper=<*Global`AreaConductor`Private`bUpper*>;
}

/**************************************
Potentialgesetz:
**************************************/

float law(float r)
{
  return <*Global`AreaConductor`Private`ForceLaw
          /. Global`AreaConductor`Private`r->r*>;
```

```
}

//////////////////////////////////////////////////////////////////////

void Force(int max_index, int selection, float *f) {
    float F, r[3], R;
    float current[3];
    int i=1-(selection>0), j;

    for(j=0;j<3;j++){ current[j]=pos3D[selection][j]; }
    f[0]=f[1]=f[2]=0.;
    for(;i<max_index;i++){
for(j=0;j<3;j++){ r[j]=current[j]-pos3D[i][j]; }
R=sqrt(r[0]*r[0]+r[1]*r[1]+r[2]*r[2]);
F=law(R);
for(j=0;j<3;j++){f[j]+=F*r[j]/R;}
if((i+1)==selection) { i++; }
    }
}

//////////////////////////////////////////////////////////////////////

void Norm(float *vector){
    int i=0;
    float length=sqrt(vector[0]*vector[0]+vector[1]*vector[1]+vector[2]*vector[2]);
    for(;i<3;i++){vector[i]/=length;}
}

//////////////////////////////////////////////////////////////////////

float Scalar(float *a, float *b){
    return a[0]*b[0]+a[1]*b[1]+a[2]*b[2];
}

//////////////////////////////////////////////////////////////////////

void Move (int max_index, int number_of_steps, int step){
    int i=0, j;
    float Factor=multiplicator*( 1.-0.5/(float)number_of_steps *
                  (float)step )/sqrt(sqrt((float)max_index)) ;
    float av[3], bv[3], F[3], a, b, shorten, dummy;

    for(;i<max_index;i++){
a=pos2D[i][0];b=pos2D[i][1];
Force(max_index, i, F);
get_tangent_vectors(av, bv, pos2D[i][0], pos2D[i][1]);
Norm(av);Norm(bv);shorten=1.;
```

```
//restrict to parameter-area:
a=Scalar(av, F);
if      ( (pos2D[i][0]+a) > a1 ){shorten=(a1-pos2D[i][0])/a;}
else if( (pos2D[i][0]+a) < a0 ){shorten=(a0-pos2D[i][0])/a;}
b=Scalar(bv, F);
if( (pos2D[i][1]+b) > b1 ){
    dummy=(b1-pos2D[i][1])/b;
    if(dummy<shorten){shorten=dummy;}
}
else if( (pos2D[i][1]+b) < b0 ){
    dummy=(b0-pos2D[i][1])/b;
    if(dummy<shorten){shorten=dummy;}
}

pos2D[i][0]+=a*shorten*Factor;
pos2D[i][1]+=b*shorten*Factor;
    }
}

//////////////////////////////////////////////////////////////////

void ChargePos(int index_max, int number_of_steps)
{
  int charge_index=0, i,j;
  double dummy3D[index_max][3],  dummy2D[index_max][2];
  long   dims3D[]={index_max,3}, dims2D[]={index_max,2};

//iterate equilibrium:
  for(i=0;i<number_of_steps;i++){
      Move(index_max, number_of_steps, i);
      for(charge_index=0;charge_index<index_max;charge_index++){
  pos3D[charge_index][0]=xParam(pos2D[charge_index][0], pos2D[charge_index][1]);
  pos3D[charge_index][1]=yParam(pos2D[charge_index][0], pos2D[charge_index][1]);
  pos3D[charge_index][2]=zParam(pos2D[charge_index][0], pos2D[charge_index][1]);
      }
  }

//check positions on area:
  for(charge_index=0;charge_index<index_max;charge_index++){
    if(pos2D[charge_index][0]<a0){pos2D[charge_index][0]=a0;}
    if(pos2D[charge_index][0]>a1){pos2D[charge_index][0]=a1;}
    if(pos2D[charge_index][1]<b0){pos2D[charge_index][1]=b0;}
    if(pos2D[charge_index][1]>b1){pos2D[charge_index][1]=b1;}
    pos3D[charge_index][0]=xParam(pos2D[charge_index][0], pos2D[charge_index][1]);
    pos3D[charge_index][1]=yParam(pos2D[charge_index][0], pos2D[charge_index][1]);
    pos3D[charge_index][2]=zParam(pos2D[charge_index][0], pos2D[charge_index][1]);
  }
```

```
//Create List for the 3D- and 2D-positions:
  MLPutFunction(stdlink, "List",2);

//"konvert" 3D-list to double for Mathematica-interface and "send" it:
  for(i=0;i<index_max;i++){
      dummy3D[i][0]=(double)pos3D[i][0];
      dummy3D[i][1]=(double)pos3D[i][1];
      dummy3D[i][2]=(double)pos3D[i][2];
  }
  charges_pointer=(double*)dummy3D;
  MLPutRealArray(stdlink, charges_pointer, dims3D, NULL, 2);

//"konvert" 2D-list to double for Mathematica-interface and "send" it:
  for(i=0;i<index_max;i++){
      dummy2D[i][0]=(double)pos2D[i][0];
      dummy2D[i][1]=(double)pos2D[i][1];
  }
  charges_pointer=(double*)dummy2D;
  MLPutRealArray(stdlink, charges_pointer, dims2D, NULL, 2);

  return;
}

////////////////////////////////////////////////////////////////////

void PosFromArray(double *RealList, long listLength,
                  int number_of_steps, double multipl){

    int i=0, charge_index, max_index=listLength/2;

    multiplicator=multipl;
    for(charge_index=0;charge_index<max_index;charge_index++){
pos2D[charge_index][0]=RealList[i];
i++;
pos2D[charge_index][1]=RealList[i];
i++;
pos3D[charge_index][0]=xParam(pos2D[charge_index][0], pos2D[charge_index][1]);
pos3D[charge_index][1]=yParam(pos2D[charge_index][0], pos2D[charge_index][1]);
pos3D[charge_index][2]=zParam(pos2D[charge_index][0], pos2D[charge_index][1]);
    }

    ChargePos(max_index, number_of_steps);
}

////////////////////////////////////////////////////////////////////

void PosFromNumber(int number_of_charges_one, int number_of_charges_two,
                   int number_of_charges_last, int number_of_steps, double multipl){
```

```
    int index_max=number_of_charges_one*number_of_charges_two+number_of_charges_last;
    int charge_index=0, i, j;

    multiplicator=multipl;
  da=(a1-a0)/(float)(number_of_charges_one+(number_of_charges_last>0));
  db=(b1-b0)/(float)number_of_charges_two;

//initialise charge-positions:
  for(i=0;i<number_of_charges_one;i++){
      for(j=0;j<number_of_charges_two;j++){
  pos2D[charge_index][0]=a0+da/2.+(float)i*da;
  pos2D[charge_index][1]=b0+db/2.+(float)j*db;
  pos3D[charge_index][0]=xParam(pos2D[charge_index][0], pos2D[charge_index][1]);
  pos3D[charge_index][1]=yParam(pos2D[charge_index][0], pos2D[charge_index][1]);
  pos3D[charge_index][2]=zParam(pos2D[charge_index][0], pos2D[charge_index][1]);
  charge_index++;
      }
  }
  for(i=0;i<number_of_charges_last;i++){
      pos2D[charge_index][0]=a0+da/2.+(float)number_of_charges_one*da;
      pos2D[charge_index][1]=b0+db/2.+(float)i*db;
      pos3D[charge_index][0]=xParam(pos2D[charge_index][0], pos2D[charge_index][1]);
      pos3D[charge_index][1]=yParam(pos2D[charge_index][0], pos2D[charge_index][1]);
      pos3D[charge_index][2]=zParam(pos2D[charge_index][0], pos2D[charge_index][1]);
      charge_index++;
  }

    ChargePos(index_max, number_of_steps);
}

//////////////////////////////////////////////////////////////////////

//////////////////////////////////////////////////////////////////////

int main(int argc, char *argv[])
{
  SetParams(&a0, &a1, &b0, &b1);
//  da=(a1-a0)/(float)AreaGrid; db=(b1-b0)/(float)AreaGrid;
  return MLMain(argc, argv);
}
```